# Strong light-matter interactions in hybrid polaritonic systems

Ben Johns,[a,b] Andrea Schirato,[c,d] Federico Toffoletti,[e] Tommaso Giovannini,[f] Mirko Vanzan,[g] Margherita Marsili,[h] Giovanni Parolin,[e] Giulia Dall'Osto,[i] Ajay Kumar Poonia,[a,j] Chiara Cappelli,[k] Francesca Baletto,[g] Stefano Corni,[e,l] Elisabetta Collini,[e] Margherita Maiuri,[c] and Nicolò Maccaferri[a,b,j]

Strong light-matter coupling gives rise to polaritons – hybrid excitations whose mixed photonic and matter character enables control over optical, electronic and chemical properties. This Feature Article surveys the main architectures supporting polariton formation, including photonic microcavities, plasmonic nanostructures, open cavities and metasurfaces, and outlines how inorganic semiconductors, organic aggregates and hybrid systems access strong and ultrastrong coupling. Key phenomena such as coherent dynamics, vibronic interactions, dark-state reservoirs and polariton-mediated energy and electron transport are discussed, together with the experimental and theoretical tools used to study them. We highlight examples where strong coupling modifies charge transport, energy flow and chemical reactivity, and we summarize emerging regimes, including intermediate and dark-strong coupling, that broaden the landscape of hybrid light-matter physics.

## 1. Introduction

Polaritons are hybrid light-matter quasiparticles formed when excitations in matter interact coherently with an optical mode. They provide a versatile platform to reshape optical, electronic and chemical properties by engineering the surrounding electromagnetic environment.[1,2] When the interaction strength between matter and light exceeds the relevant dissipation rates (lifetimes), the system enters the so-called strong coupling regime.[3] Here, material excitations and photons no longer behave as separate entities, but form mixed eigenstates with new energy landscapes and, possibly, new dynamical pathways.

A defining signature of this regime is the Rabi splitting, which appears as a consequence of the anti-crossing between the bare material excitation and the dispersion of the optical mode. Its magnitude reflects the coherent energy exchange rate between "matter and light" and determines the composition, coherence and lifetime of the resulting upper and lower polariton branches (UP and LP, respectively). In frequency-domain measurements, this splitting provides direct evidence of hybridisation, while time-resolved spectroscopy reveals corresponding Rabi oscillations, which are a direct proof of ultrafast population beatings that embody the light-matter coupling in the time domain. By using different architectures, such as high quality-factor (Q) photonic microcavities, plasmonic nanostructures offering extreme mode confinement, and open cavities providing tailored optical modes, the Rabi splitting and polaritons dynamics can be tuned and manipulated.

This Feature Article presents a unified perspective on hybrid polaritonic systems where electronic excitations are coupled to optical cavities, beginning with the fundamental building blocks that enable strong coupling (section **1.1**). We discuss photonic microcavities, plexcitonic platforms based on metallic nanostructures, and emerging open-cavity and metasurface architectures (**Figure 1**, top left). We further emphasize how the choice of the materials with specific electronic excitations, from inorganic Wannier-Mott excitons to organic Frenkel excitons such as J-aggregates, has expanded access to strong and ultrastrong coupling regimes under ambient conditions.

We then explore the rich physical mechanisms that arise from strong coupling in section **1.2**, including coherent dynamics, vibronic-polaritonic interactions, polariton transport, and the interplay between bright polariton branches and dark-state reservoirs (**Figure 1**, top right). Particular attention is devoted to polariton-enabled energy and electron transport, where hybridisation affords ultrafast and long-range propagation pathways inaccessible to uncoupled materials.

To investigate these effects, we examine the experimental methodologies used in polaritonics (**Figure 1,** bottom left), such as Fourier microscopy, transient absorption spectroscopy, and two-dimensional electronic spectroscopy (section **2.1**), alongside state-of-the-art theoretical methods, ranging from continuum electrodynamics to multiscale quantum/classical descriptions capable of resolving molecular detail and pico-scale plasmonic response (section **2.2**).

a. Ultrafast Nanoscience Group, Department of Physics, Umeå University, 90187 Umeå, Sweden. Email: nicolo.maccaferri@umu.se
b. Umeå Centre for Microbial Research, 90187, Umeå, Sweden
c. Dipartimento di Fisica, Politecnico di Milano, Piazza Leonardo da Vinci 32, 20133 Milan, Italy.
d. Department of Physics and Astronomy, Rice University, 6100 Main Street, Houston, TX 77005, United States
e. Dept. of Chemical Sciences, University of Padova, via Marzolo 1, 35131 Padova, Italy
f. Department of Physics, University of Rome Tor Vergata, Rome, Italy
g. Department of Physics, University of Milan, Via Celoria 16, 20133 Milan, Italy
h. Dipartimento di Fisica e Astronomia "Augusto Righi", University of Bologna, Viale Berti Pichat 6/2, 40127 Bologna, Italy
i. Elettra Sincrotrone Trieste, SS 14 Km 163.5 in AREA Science Park, Basovizza, Trieste, Italy
j. Wallenberg Initiative Materials Science for Sustainability, Department of Physics, Umeå University, 90187 Umeå, Sweden
k. Scuola Normale Superiore, Piazza dei Cavalieri 7, I-56126 Pisa (Italy)
l. Istituto Nanoscienze – CNR, S3, Via G. Campi 213/A, 41125 Modena, Italy

Finally, in section **3** the article surveys emerging applications (**Figure 1**, bottom right), including polariton-enhanced optoelectronic devices for long-range energy and charge transport, and coupling regimes such as dark-strong coupling that challenge conventional definitions of strong coupling. We conclude this article by highlighting open challenges, including the treatment of vibronic structure, hot-carrier dynamics, chiral and collective effects, and predictive modelling of complex nanosystems, and outlining future directions we believe will be the driver for new insights into the science of polaritons.

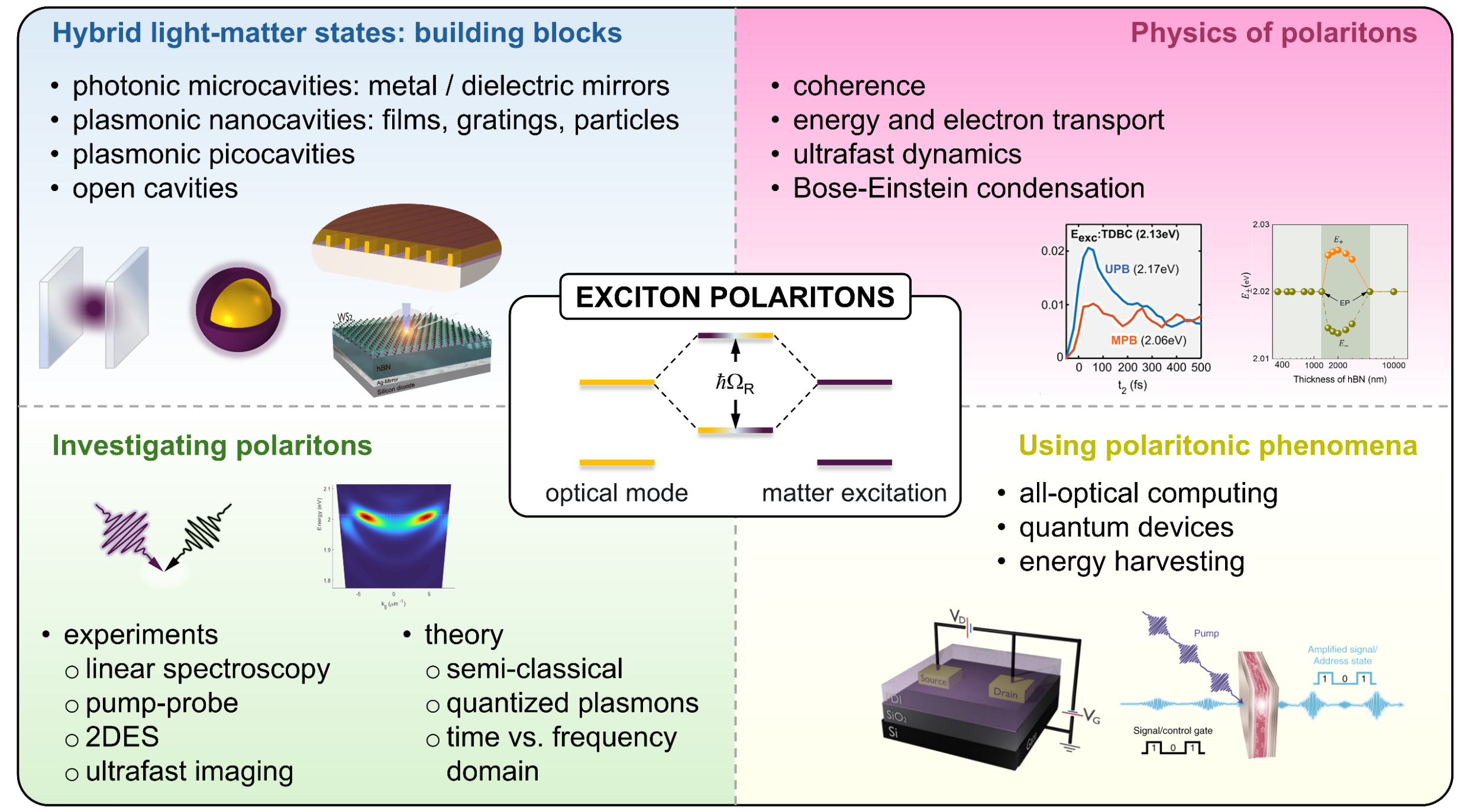

**Figure 1**. Conceptual overview of hybrid polaritonic systems and their key ingredients. The figure summarises representative photonic and plasmonic building blocks enabling strong light-matter coupling (top left), key physical mechanisms emerging from polariton formation (top right), experimental and theoretical approaches employed to investigate polaritons (bottom left), and selected application areas enabled by polaritonic effects (bottom right). Clockwise from top left: reproduced with permission from ref.[4], Copyright 2025 Wiley-VCH GmbH; ref. [5] Copyright 2024 The Author(s), licensed under a CC-BY Creative Commons Attribution 4.0 License; ref.[4] Copyright 2025 Wiley-VCH GmbH.; ref.[6], Copyright 2019, The Author(s), under exclusive licence to Springer Nature Limited; ref.[7], Copyright 2025, American Chemical Society; ref.[4], Copyright 2025 Wiley-VCH GmbH.

### 1.1 Hybrid light-matter states and their building blocks

In this section, we introduce three classes of hybrid systems where strong light-matter interactions have been realized: (i) photonic microcavities, (ii) plexcitonic systems and (iii) 'open' cavities (**Figure 1**, top-left panel). The first category represents a well-established and versatile platform for investigating polaritonic phenomena, enabling fundamental studies in fields such as polaritonic chemistry[8] and nonlinear polaritonics.[9] In parallel, plexciton architectures, where excitons strongly couple to surface plasmon resonances, have attracted significant attention, owing to their potential to drive chemical and quantum processes at sub-wavelength length scales.[10] 'Open' cavity architectures have more recently emerged, offering advantages such as flexible tuning of modes and ease of integration of a wide range of materials, including layered semiconductors and organic molecules.[7]

#### 1.1.1 Photonic microcavities

Photonic Fabry-Pérot microcavities, based on both metallic and dielectric mirrors, have been extensively employed to access strong coupling with excitons. They offer significant advantages, including relatively high Q-factors (10-100) of the optical modes,[11] and mature fabrication methods. On one hand, dielectric mirrors offer lower losses (and hence, higher Q-factors), which however comes at the price of lower mode confinement.[12] On the other hand, metallic mirrors ensure better confinement, but with higher losses. More generally, photonic cavities have been the test bed to explore the physics of polaritons under controlled conditions. Light-matter coupling between epitaxially grown Wannier-Mott excitonic layers and semiconductor microcavities was thoroughly explored in the last century, when the fundamental properties of strongly coupled systems were established.[13] The drawback of these materials is their low exciton binding energy, which necessitates performing the experiments under cryogenic conditions. To overcome this limitation, materials with larger binding energy have been explored more recently, such as two-dimensional (2D) transition metal dichalcogenides (TMDCs) and metal halide perovskites. 2D materials are exciting for many reasons, such as tunable bandgaps and strong photoluminescence at visible wavelengths. They are also an ideal platform to explore many-body phenomena (trions, bi-excitons) and valley-selective properties due to their strong spin-orbit coupling.[14] Exciton-polaritons formed by hybridizing TMDC monolayers with various cavity architectures have been reported, including localized plasmons,[15] surface plasmon polaritons,[16] Fabry-Pérot cavities[17,18] and thin film cavities.[4] In parallel, metal halide perovskites have emerged as an attractive platform owing to their ease of processing and tunable bandgaps, demonstrating room temperature polariton formation and condensation in Fabry-Pérot microcavities.[19] Finally, organic dyes and semiconductors constitute another promising class of excitonic materials, which have been explored for room-temperature strong coupling. Their large oscillator strengths have enabled access to the ultrastrong coupling regime[20] and provide the flexibility of engineering structure and morphology (aggregates, supramolecular assembly, etc.). Moreover, their fabrication is typically simpler (evaporation or solution processing) than conventional inorganic semiconductors, which often require epitaxial growth. Research on strong coupling with organic molecules has given rise to the field of molecular polaritonics,[21] which shows tremendous potential in controlling chemical reactions and modifying quantum processes.[22,23]

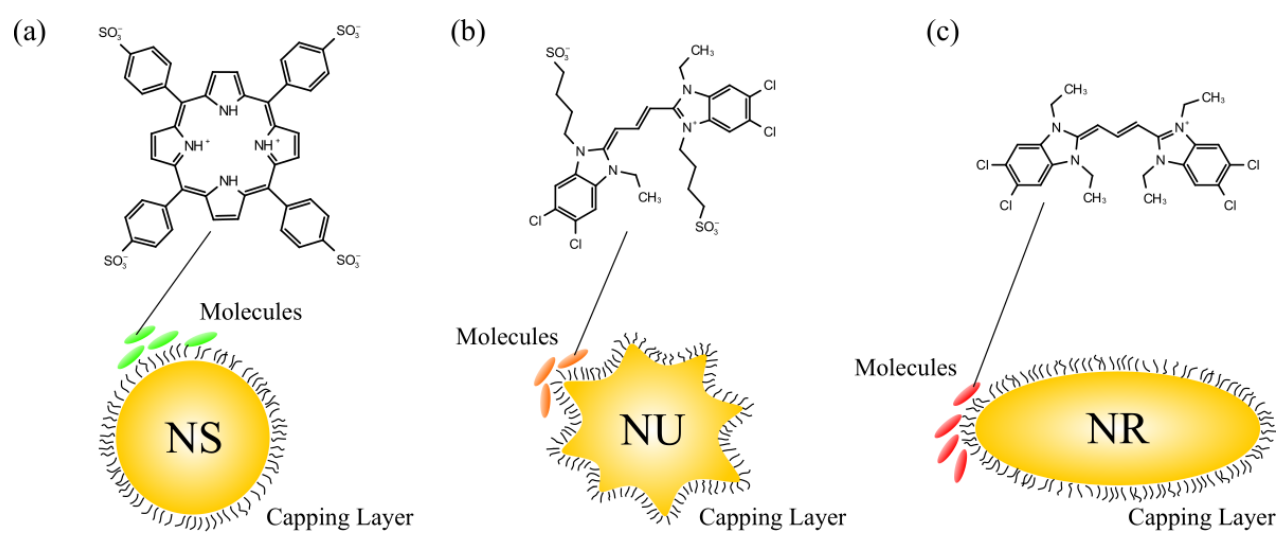

**Figure 2**. Schematic representation of selected examples of plexcitonic colloidal nanohybrids. (a) Gold nanospheres functionalized with tetra(sulfonatophenyl)porphyrin (TPPS) (ref. [24,25]) (b) gold nanourchins functionalized with tetrachloro-diethyl-di(4-sulfobutyl)benzimidazole carbocyanine (TDBC) J-aggregates [ref.[26]]; (c) gold nanorods functionalized with with 5,5′,6,6′-Tetrachloro-1,1′,3,3′-tetraethylbenzimidazolylcarbocyanine (JC-1) Iodide [ref.[27]]

#### 1.1.2 Plexcitons – molecular emitters coupled to plasmonic nanoparticles (NPs)

Plasmonic nanostructures offer many advantages over photonic cavities, including smaller mode volume and stronger field enhancements, albeit at the cost of lower Q-factors.[28] Although plasmonic systems exhibit higher dissipation rates (due to the low Q-factor $\leq 10$) and hence shorter lifetimes (typically only a few tens of femtoseconds[29,30]), they can still achieve strong coupling thanks to the exceptional field enhancement within small mode volumes, of the order of $1-100$ nm$^3$.[31] The strong confinement of plasmonic modes is considered crucial for reaching the quantum strong coupling regime with single emitters or molecules.[32] Plexcitonic strong coupling has been explored in various cavity geometries involving both propagating and localized surface plasmons. In the case of propagating plasmons (surface plasmon polaritons), strong coupling has been investigated on metallic films (holes,[33] arrays,[34] slits[35] etc.). For localized surface plasmons, nanostructures including nanospheres, rods, cubes, core-shell structures, and others, have been used (**Figure 2**).[36,37] For even smaller mode volumes ($V_m \sim 10^{-5}\,\lambda^3$, compared to $V_m \sim 10^{5}\,\lambda^3$ in typical photonic microcavities[38]) and larger field enhancements, nanocavities of plasmonic dimers and bowties are employed. A particularly promising platform for enhancing electromagnetic field localization is the nanoparticle-on-mirror configuration, consisting of a nanostructure placed at a small (typically sub-nm) distance above a metallic film and supporting a gap plasmon,[39] which allows fine control over plasmon energy and extreme light confinement.[40]

On the matter side, plexcitons have been realised using a variety of material systems,[14] such as molecular dyes and J-aggregates,[24–26,35] organic photo switches,[31] and metal halide perovskites.[19] These hybrid architectures based on metallic nanostructures have favourable properties such as scalability, spectral tunability and ease of synthesizing through wet chemistry. Moreover, colloidal, organic-based plexcitons can operate at room temperature[26] and can be selectively activated or tuned by manipulating external conditions, e.g. dye concentration.[24,41]

#### 1.1.3 Open cavities

An emerging direction in strong coupling research is the coupling of excitons to 'open' optical cavities. The geometry of conventional microcavity structures introduce limitations on the ability of molecules to interact with an external environment.[42] To address these difficulties in fields such as cavity-modified chemical reactivity, where access to the cavity-coupled molecules is highly desired, alternate cavity geometries have been explored. In the context of strong coupling, the term open cavity has been used to refer to either plasmonic cavities, or planar cavities supporting modes formed by the refractive index contrast of its layers. Recent research efforts have shown that strong coupling can be realized in cavities with one or even no mirrors, significantly expanding the palette of optical platforms that support polariton formation.[7,17,43,44] Similarly, self-hybridized exciton-polaritons have also been explored where photonic modes supported by the material geometry itself hybridizes with the excitons.[45–47] Finally, a cavity architecture that has recently gained momentum, positioned at the intersection of conventional photonic microcavities and plasmonic colloidal NPs, employs resonant optical metasurfaces. These quasi-2D ordered arrays of nanoantennas provide unparalleled degrees of freedom to shape and enhance local fields at the deep sub-wavelength scale, and to finely tailor near-field interactions.[48] In the context of strong coupling, both metallic and all-dielectric, local and nonlocal metasurfaces acting as optical resonator offer unique advantages, in particular ease of in-/out-coupling, owing to their open cavity character and virtually unlimited tunability of resonance spectral position, modal profile and lifetime (i.e., Q-factor).[49,50]

### 1.2 The intriguing physics of polaritons

#### 1.2.1 Coherence

The key dynamical signature of the strong coupling regime is the appearance of Rabi oscillations, i.e., a coherent and reversible exchange of energy between the transition dipole of the material and the vacuum electromagnetic field of the cavity. Originally described in atomic and semiconductor systems, Rabi oscillations manifest themselves as a periodic population oscillation with a frequency determined by the coupling strength, and they provide the microscopic origin of the characteristic normal mode splitting in the polariton dispersion in the frequency domain.[51] This ability to sustain coherence makes polaritons a unique platform to study coherent dynamics and interactions.[24–27]

***Rabi oscillations in microcavity polaritons.*** The emergence of cavity polariton physics can be traced back to the pioneering observation by Weisbuch and coworkers, reporting Rabi splitting in a Fabry-Pérot microcavity embedding GaAs quantum wells.[52] This steady-state measurement provided the first unambiguous evidence of the strong-coupling regime between excitons and cavity modes. Soon after, Norris et *al.*, reported the first time-resolved ultrafast optical measurement of Rabi oscillations in such a microcavity system, using femtosecond pump-probe spectroscopy to directly track the coherent exciton-photon population exchange in the time domain.[53] More recently, the coherent control of polariton Rabi oscillations at the femtosecond scale was demonstrated by Dominici, *et al.* in a semiconductor microcavity,[54] where ultrafast pulses were used to manipulate the amplitude and phase of light-matter oscillations. These studies established the experimental basis for observing, resolving, and controlling Rabi oscillations in Fabry-Pérot microcavities.

***Rabi oscillations in plexcitons.*** The first direct observation of Rabi oscillations in plasmonic cavities was reported by Vasa et al. in 2013 using femtosecond pump-probe spectroscopy (**Figure 3a**).[35] In their pioneering experiment, J-aggregates of cyanine dyes were deposited on a gold nanoslit array, forming hybrid polaritonic states via strong coupling with propagating surface plasmon polaritons. This coupling was evidenced by normal mode splitting and, crucially, by temporal oscillations in the pump-probe signal, reflecting coherent energy exchange between the excitons and surface plasmon polaritons. These Rabi oscillations, with a period of ~30 fs and lasting ~50 fs, could be actively modulated by varying the exciton density on a 10-fs timescale. Despite involving only a small excitonic fraction, the experiment clearly demonstrated ultrafast coherent dynamics.
Coherent Rabi oscillations were also observed in solution-phase plexcitonic nanohybrids composed of gold nanourchins functionalized with J-aggregates (**Figure 3b**). Using two-dimensional electronic spectroscopy (2DES, see **section 2.1.2**), Peruffo et al. identified diagonal and cross-peaks associated with upper and lower plexciton states.[26] Oscillatory signals, persisting on femtosecond timescales (~16 fs), confirmed coherent light-matter interaction even in disordered systems at room temperature.

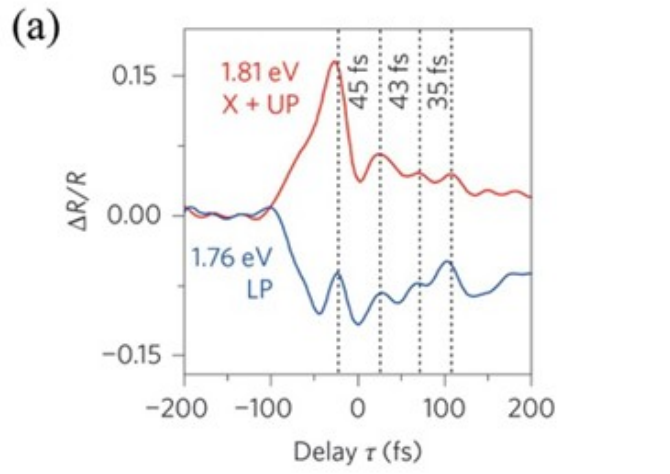

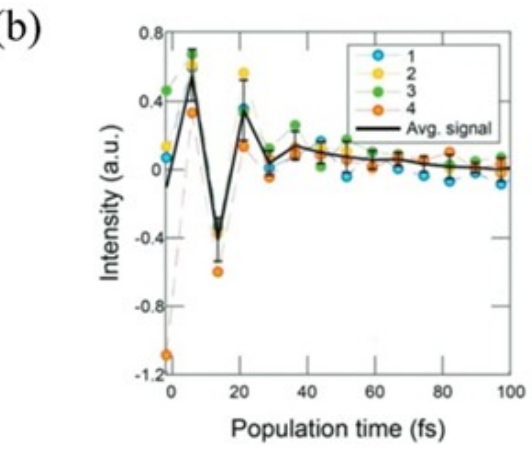

Figure 3. Evidence of Rabi oscillations in plexcitonic materials. (a) Time dynamics of the transient reflectivity for a system composed of J-aggregates of cyanine dyes deposited on a gold nanoslit array. The oscillatory behavior, with periodicities corresponding to the Rabi oscillation, is shown as a function of pump-probe delay (τ). Reproduced with permission from ref.[35], Copyright 2013 Springer Nature. (b) Signal decays (intensity vs time) at relevant coordinates extracted from the 2DES response of nanohybrids composed of gold nanourchins functionalized with J-aggregates. Coherent beatings are reproducibly found with a dephasing time of about 16 fs. Reproduced with permission from ref.[26] Copyright 2023 The Author(s), licensed under a Creative Commons CC-BY-NC-ND license.

***Effect of vibronic polariton states on coherence***. An emerging area of interest concerns the interplay between polaritonic and vibronic coupling, which gives rise to vibronic-polariton states. The properties of these states are currently attracting considerable attention due to their potentially decisive role in shaping coherent dynamics, and in the stabilization and control of polaritonic systems.[55] Vibronic coupling, arising from the interaction between electronic and vibrational degrees of freedom, is a defining feature of organic and hybrid materials. Beyond acting as a source of disorder, molecular vibrations can strongly couple to electronic transitions and profoundly shape their dynamics. When the light-matter interaction becomes comparable to the vibronic coupling, as described by the Holstein-Tavis-Cummings model,[56,57] different regimes of hybrid vibronic polaritons emerge. In this context, phenomena such as polaron decoupling—where strong Rabi coupling suppresses vibronic interactions and preserves Rabi oscillations—occur when the Rabi splitting exceeds the vibronic coupling.[58–60]
It is important to emphasize that, beyond being a source of decoherence, vibronic coupling can also sustain coherence and mediate population exchange between polaritonic, excitonic, and dark states. Theoretical and experimental studies have shown that vibronic coupling can actively promote coherence and bright-dark (or bright-gray) interconversion,[61,62] in analogy with purely excitonic systems.[63] In addition, recent quantum-dynamical simulations have highlighted vibronic coupling as a key ingredient in polariton transport.[64] In particular, vibrational interactions were shown to mediate efficient population exchange between the slow-propagating exciton dark-states and the fast-propagating polaritons sustaining ballistic transport. In particular, ref.[27] experimentally demonstrated that molecular vibrational modes can act as resonant channels enabling coherent interactions between the lowest polariton level and partially dark-states, providing direct evidence that vibronic coupling has a central role in shaping the coherent dynamics in disordered plexcitonic systems.

### 1.2.2 Energy and electron transport

The photonic component of polaritons can sustain transport phenomena far beyond the limits of their excitonic constituents, enabling enhanced charge transport and long-range, ultrafast energy propagation.[65,66] While bare excitons in organic materials diffuse only over nanometers due to the inherent tightly bound nature and their highly disordered character,[67] polaritons have been directly imaged propagating ballistically across microns within a few hundred femtoseconds, with velocities of tens of microns per picosecond.[68] This ballistic behaviour has been demonstrated to persist up to ~50% excitonic fraction, where hybridization shields the wave packet from dynamic disorder, before strong exciton-phonon interactions drive decoherence and diffusive transport.[68,69] Such delocalization underpins polariton-mediated energy transfer across molecular ensembles in microcavities, where hybrid states enable energy flow over hundreds of nanometres to microns.[70–73]

A prototypical system for energy transfer consists of two different molecular states that are hybridized with a single cavity mode. The molecular states are spatially separated well beyond conventional non-radiative energy transfer distances, nonetheless the excitation energy of a 'donor' can still be efficiently transferred to an 'acceptor' molecule. The extent of delocalization of photoexcitation can be very large (microns) relative to atomic length scales (see **section 3.2.1**).[5,23,74,75] When an exciton hybridizes with an optical mode that has large spatially propagating character (e.g., propagating plasmon, Bloch modes), the polariton wavefunction is delocalized over many molecules. This can result in modified charge transport as well.[33,76] In high-Q dielectric architectures supporting e.g. Bloch surface waves, polariton lifetimes extend to a few hundred femtoseconds—sufficient to compete with nuclear reorganization times—allowing direct observation of polariton-assisted charge transfer with reduced driving force and high quantum efficiency. [77] On the other hand, hybridization with a cavity mode also allows enhancement of charge transport under strong coupling,[78–80] even with low-Q resonances such as those in thin-film or one-mirror cavities, which could enable efficient electronic devices driven by polaritonic effects (see **section 3.2.2**). These results have shown that strong coupling enables ultrafast, long-range energy transfer as well as the reshaping of electronic conduction, providing new pathways for photocurrent generation and polaritonic optoelectronics. The demonstration of enhanced energy and electron transport processes has the potential to improve organic photovoltaics, light sources, lasers, and optoelectronic devices such as cavity-coupled phototransistors.[81,82]

### 1.2.3 Polariton dynamics and dark states

From the viewpoint of dynamics, the formation of polaritons modifies the system's energy landscape and introduces new pathways for energy relaxation.[23,55,66] Polariton dynamics is particularly rich, as these hybrid states emerge from the interplay of coherent light-matter exchange and multiple incoherent relaxation channels. Indeed, the bright lower and upper polariton branches coexist with a dense manifold of dark reservoir states. Moreover, polaritons display unique coherent dynamics that are observed as oscillatory signals (described in **section 1.2.1**), which are hampered by incoherent relaxation processes such as plasmon dephasing or phonon scattering. This presents the need for a detailed understanding of interaction and relaxation pathways to rationalise polariton dynamics in detail. For a long time, this "dark-state problem" has limited our ability to unambiguously assign lifetimes and relaxation channels, as broadband femtosecond excitation typically populates the reservoir and only indirectly modifies the polariton response.[83] As a result, early ultrafast experiments often captured dispersive signals linked to exciton bleaching or Rabi splitting renormalization, rather than the intrinsic polariton population. Nevertheless, dark states may also serve as a valuable resource for manipulating the dynamical properties of strongly coupled materials. It has been shown, for instance, that the ultrafast dynamics of plexcitonic nanohybrids are profoundly influenced by the presence of a state reservoir—including plasmonic resonances and dark states—and can be tuned by adjusting the excitation wavelength and fluence. By varying the number of interacting plexciton resonances (thus altering the distribution of reservoir states) and modulating the excitation fluence, researchers observed relaxation dynamics spanning at least an order of magnitude in characteristic timescales.[25] This finding opens an intriguing perspective: leveraging strong light-matter coupling to actively tailor the rates of molecular processes and chemically relevant reactions.

Recent advances have begun to resolve the ambiguity related to dark-state dynamics. By exploiting momentum-resolved, phase-matched excitation, experiments can now directly populate the lower polariton branch and follow its evolution in time, revealing sub-100 fs lifetimes and ultrafast momentum scattering distinct from the slower dynamics of reservoir states.[83] These measurements also highlight the key role of exciton-photon composition in setting polariton lifetimes, with more excitonic states exhibiting longer-lived dynamics (even an order of magnitude, e.g., 0.1–10 ps in other systems vs >10 ps in plexcitons compared to other polaritonic systems, aided by the synergistic action of the plasmonic band and dark states).[25]

# 2 Theoretical and Experimental Methods

## 2.1 Experimental methods to investigate light-matter interactions in photonic/plasmonic hybrid systems

### 2.1.1 Linear spectroscopy and microscopy

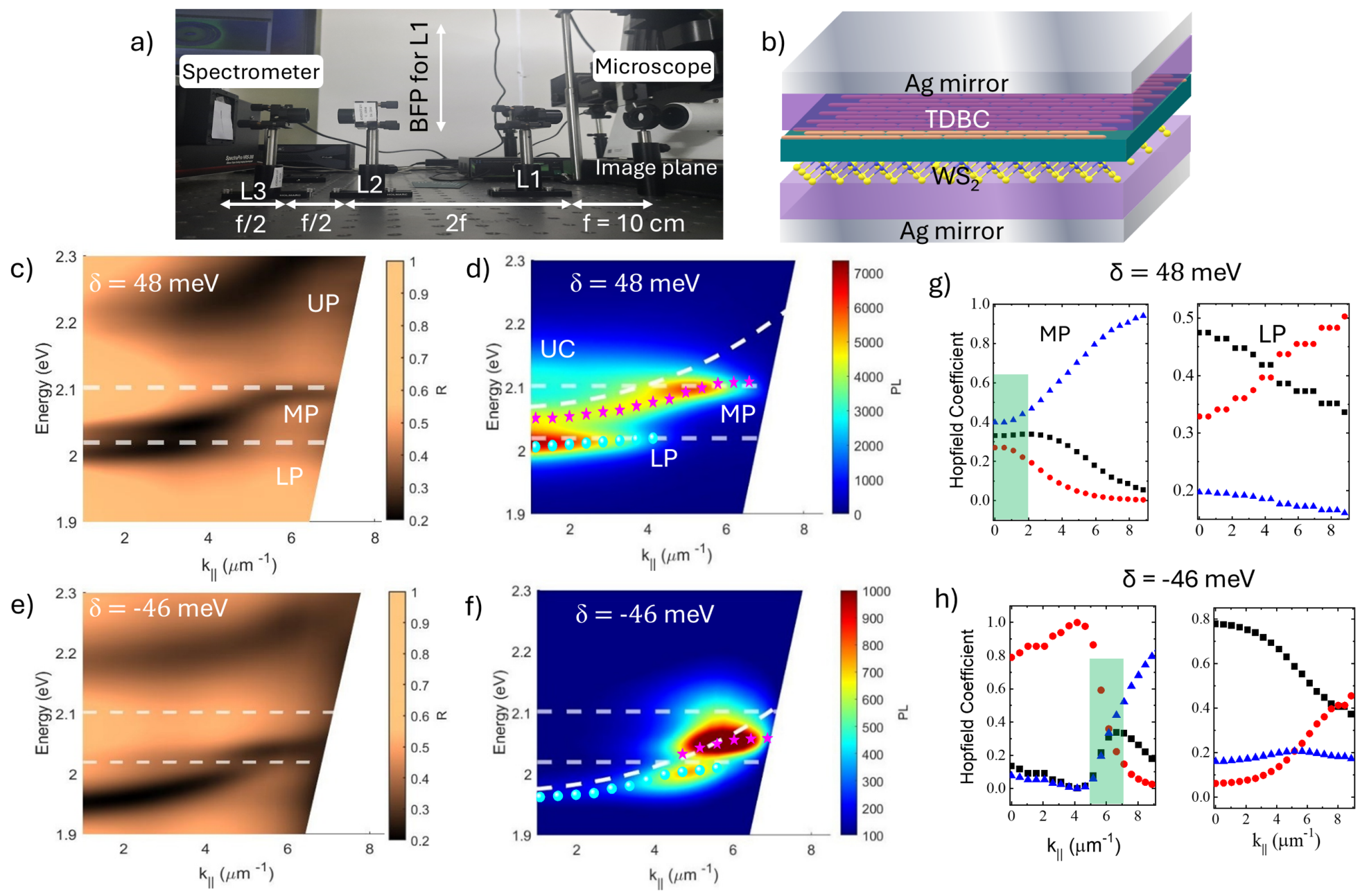

**Figure 4**. Fourier microscopy and spectroscopy of an organic-inorganic hybrid polariton system. (a) Schematic of Fourier microscopy setup, (b) organic-inorganic hybrid system consisting of a $WS_2$ monolayer and thin film of TDBC, (c,e) measured dispersion in reflectance mode and (d,f) PL mode using Fourier microscopy for positive and negative exciton-cavity detuning $\delta$. Estimation of Hopfield coefficients based on dispersion measurements for (g) positive and (h) negative detuning. Adapted with permission from ref. [84] Copyright 2025 Wiley-VCH GmbH.

The conventional approach to probe microcavity polaritons is to measure simultaneously the energy spectrum and the angular distribution of light exiting the cavity. In this way, one can directly observe the polariton dispersion, resulting from the coupling of the cavity mode and an excitonic transition, and extract the energy of their coupling strength (i.e., the Rabi splitting). In typical planar photonic cavities such as semiconductor microcavities, Fabry-Pérot cavities, and thin film cavities, the dispersion can be measured straightforwardly from spectral measurements at varying angles of incidence of light.[85,86] However, when the lateral dimensions are of the order of micrometres or less, the experimental method commonly used is Fourier microscopy, also referred to as k-space imaging.[87,88] Light transmitted or reflected from the cavity is collected by a series of lenses in a 4-f-configuration, which uses a Fourier lens to perform a spatial Fourier transform of the image plane, in which the spatial coordinates correspond to the (*x,y*) components of the emitted/transmitted light wavevector.[89] **Figure 4a** schematically shows a setup for Fourier microscopy combined with a spectrometer.[4,7,18,78,84] By imaging the Fourier plane onto a spectrometer coupled with a CCD, one can directly measure the angular distribution of light as a function of wavelength.[17,90–92] In **Figure 4a**, a 100x microscope objective (0.75 N.A., 0-50º angular range) focuses light onto the sample, and collects the reflected light. Two achromatic lenses (f = 100 mm) relay the image plane onto the CCD array of the spectrometer camera. Introducing an additional achromatic lens (f = 50 mm) into the optical path brings the Fourier plane at the position of the CCD. The desired polarization (TE or TM) can be selected by placing a polarizer between the two f = 100 mm relay lenses in the 4-f configuration.

Fourier microscopy has been employed effectively to map the dispersion and emission characteristics in 2D semiconductor-based polaritonic systems,[17,18,84] where typical flake sizes are of the order of 10 μm. This technique is also useful in related systems where multiple excitonic resonances (e.g. organic and 2D semiconductor excitons) couple to the same cavity mode,[93] the resulting polaritonic states (lower, middle and upper) contain varying degrees of photon and exciton components, and therefore allows one to design hybrid systems that combine the desirable properties of its constituents.[81,82,94] Dutta et al. investigated polariton emission in a hybrid organic-inorganic microcavity composed of organic Frenkel excitons in TDBC and Wannier-Mott excitons in the inorganic semiconductor $WS_2$ (**Figure 4b**) using Fourier microscopy.[84] By varying the exciton-cavity detuning ($\delta$) across positive and negative values, a straightforward control over the relative mixing is possible, as visualized in the reflection dispersion maps (**Figure 4c,e**). Using Fourier imaging to measure the photoluminescence (PL) signal as an additional way to gain insight into the photophysics of the hybrid system, Dutta et al. further showed that the energy and angular dependence of polariton emission, especially from the middle polariton, was strongly influenced by cavity detuning (**Figure 4d,f**). This was attributed to the relative light-matter mixing fractions, defined by the Hopfield coefficients, which can be extracted from the measured dispersion. The Hopfield coefficients obtained as a function of the in-plane wave vector revealed a correlation between the photonic character and increased PL emission from the middle polariton (**Figure 4g,h**). In **Section 3.2.2,** another example is presented, where Fourier microscopy was used to correlate polariton dispersion measurements with an enhancement of electron transport in a strongly coupled field effect transistor device.

For completeness, we mention that slightly different approaches are used in plasmonic and NP systems to observe the anti-crossing behaviour in the dispersion when localized optical resonances are involved. Instead of obtaining an energy-momentum dispersion, anti-crossings in the dispersion plotted as energy against the NP size or nanocavity dimension (which govern spectral detuning from the excitonic resonance) can be used to identify the Rabi splitting. [16,95,96] For this, the size of the resonator[97] or refractive index of the surrounding medium[98] are varied to shift the energy of the plasmonic cavity mode, and gradually detune it from the (momentum-independent) exciton energy.

Various authors have highlighted the need for unambiguous experimental determination of Rabi splitting. For instance, it has been noted that the Rabi splitting determined from an individual channel, such as scattering or reflection, may not represent the actual splitting of polariton energy levels.[21,99] This calls for care when interpreting, for example, only the scattering spectrum of NPs, or the reflection spectrum of planar cavities. Recently, it was pointed out that the Rabi splitting measured from optical spectra can deviate from the actual polariton eigenenergy levels (also called the level splitting).[100] This could lead to discrepancies in interpreting the light-matter coupling regime if care is not taken to accurately identify the energy levels of the hybridised system. In microcavity polaritons or plexciton systems, this mismatch between the spectrally observed Rabi splitting (SRS) and level splitting may be noticeable when the splitting is comparable to the line widths of the interacting states.[101,102] **Section 3.2.3** describes an example where the SRS and the actual polariton level splitting were found to be different, leading to the observation of the so-called dark-strong coupling regime.

#### 2.1.2 Nonlinear spectroscopy

Ultrafast nonlinear spectroscopy techniques use sequences of ultrashort (from hundreds down to a few femtoseconds) light pulses to investigate photoinduced dynamical processes in a variety of systems, from atoms and molecules to nanostructures.[103] When applied to polaritons, these experimental approaches represent a powerful tool to access, track, and possibly manipulate mechanisms such as excited state relaxation[104–108] and energy transfer.[109,110]

In the last decades, several techniques with different degrees of sophistication have been developed.[103] Among all, transient absorption (TA) spectroscopy remains the simplest, most powerful approach. TA spectroscopy works in a stroboscopic fashion, with a sequence of two pulses. A first intense 'pump' pulse initiates a dynamic photophysical process, bringing the system out of equilibrium. A second delayed, typically broadband pulse, referred to as the 'probe', experiences the perturbed system. By systematically varying the delay between pump and probe pulses using an optical delay line, the probe effectively tracks in real time the evolution of the excited system, by monitoring its third-order nonlinear optical response to extract pump-induced changes in its absorption spectrum.[111]

Despite the invaluable contributions and insights into polaritonic systems extracted from pump-probe spectroscopy, when applied to track the polariton relaxation dynamics, ultrafast measurements may become challenging to perform and interpret, primarily due to (i) the ultrashort lifetime of the cavity photons, depending on the mode Q-factor; and (ii) the ultrafast polariton decay, mainly towards the reservoir of dark states.[83,112] Investigating timescales on which polaritons are coherent, namely when the wavefunctions of excitons and photons are indistinguishably mixed, calls for pulses whose duration is shorter than the polariton lifetime. Combined with the need to better identify relaxation dynamics and coherent lifetimes, this has led to the implementation of 2DES in the study of polaritons.

2DES is an advanced ultrafast optical technique using three laser pulses to gather detailed femtosecond-resolved relaxation dynamics. Unlike pump-probe, 2DES provides a 2D map $\boldsymbol{S}^{(3)}(\boldsymbol{\omega_1}, \boldsymbol{t_2}, \boldsymbol{\omega_3})$, correlating excitation ($\boldsymbol{\omega_1}$) and detection ($\boldsymbol{\omega_3}$) frequencies as a function of waiting time ($\boldsymbol{t_2}$). This allows for distinguishing homogeneous and inhomogeneous broadening,[113–115] tracking spectral diffusion,[116,117] observing energy/charge transfer processes in spectrally congested systems,[118,119] including polaritons.[120,121] For polaritons, 2DES is crucial for disentangling complex dynamics, revealing energy transfer to dark states, and extracting quantum beating signals that show coherent dynamics between states.[26,122,123] It provides a comprehensive view of how light-matter coupling modifies material properties.[124,125] **Section 3.2.1** describes an experimental verification of long range energy transfer via polaritonic states using 2DES.

### 2.2 Theory: basics of hybrid modeling using quantum chemistry for molecules and classical continuum or atomistic for plasmonic nanostructures

#### 2.2.1 Overview of general methods for modelling nanoparticles and nanoparticle-molecule interface

A key element of plexcitonic systems is the nanostructure supporting plasmon excitations, that in many cases involves plasmonic NPs. The accurate description of plasmonic NPs, particularly when interacting with molecular adsorbates or complex environments, requires computational approaches capable of spanning multiple length and time scales. Over the years, a wide range of methods have been developed, with atomistic approaches emerging as an accurate framework to study plasmon-molecule strong coupling.

In atomistic approaches, the atomic structure is explicitly considered, capturing both the band structure and precise morphology of NPs. This level of detail allows direct investigation of how surface features, adsorption sites, and molecular orientation affect the emergence of hybrid light-matter states and field enhancement at the interface. First-principles simulations based on Density Functional Theory (DFT) and Many-Body Perturbation Theory (MBPT) are reference methods for studying charge transfer, hot-carrier relaxation, and plasmon-induced hybridization. Although computationally demanding and limited to systems of a few hundred atoms, these approaches provide valuable mechanistic insight into the photophysics and reactivity of plasmonic nanostructures[126–128], allowing direct access to the strong coupling regime.[129] To reduce computational cost and extend the accessible system size, simplified approaches have been developed, such as the polTDDFT algorithm,[130,131] which enables efficient time-dependent (TD)-DFT simulations on larger systems.

As the optoelectronic properties of plasmonic NPs depend strongly on size, shape, and surface arrangement, accurate morphological modelling is essential. Classical molecular dynamics simulations allow characterization of NP structure and assembly under realistic conditions, offering atomically resolved input for modelling plasmon-molecule coupling and spectral tuning, see e.g., the case of metallic nanoalloys.[132–139]

Classical methods that model the nanosystem as a continuous metallic body described by a frequency-dependent dielectric function are also fundamental in this field. Within this framework, plasmonic resonances are obtained by solving Maxwell's equations without explicit atomic detail. Widely used implementations include the finite-difference time-domain (FDTD), finite element method (FEM), boundary element (BEM), and discrete dipole approximation (DDA) methods. These approaches are essential for analysing mode hybridization, spectral splitting, and field confinement in the strong-coupling regime, enabling simulations of systems extending to hundreds of nanometers.[140] Classical models that include the atomistic structure have also been proposed to model nanostructures more realistically, and will be discussed later.

Recognizing the limitations of purely quantum or classical descriptions, the field increasingly relies on hybrid multiscale methods, now essential for capturing strong coupling across multiple energy and length scales. In these frameworks, quantum chemistry approaches are exploited to describe the molecular or reactive region, while classical, either continuum or atomistic, models are used to describe the NP.[141] Such strategies enable realistic simulations of NP-molecule interfaces where energy exchange, Rabi splitting, and charge-transfer dynamics are treated quantum-mechanically, while the extended electromagnetic response is captured at lower computational cost.[141–143]

It is worth remarking that the development of multiscale hybrid quantum/classical methods for strong coupling requires a proper description of the quantized plasmons, which in turn requires reformulating commonly exploited classical models within a Lagrangian and Hamiltonian formulation. All electrodynamics-based classical approaches for nanoplasmonics can be reformulated in this framework due to their common dissipative nature, as detailed in ref. [144] In the following, one such hybrid quantum/classical method (the polarizable continuum model – nanoparticle, PCM-NP, approach), that has been used for investigating various plexcitonic phenomena, will be described.

### 2.2.2 Basis of the Polarizable Continuum Model (PCM-NP) approach

Hybrid multiscale models benefit from a BEM description of the classical continuum portion. This allows a more computationally effective interface with the quantum part, since the spatial grid that controls classical-quantum interaction is 2D, as opposed to the 3D grid of FEM and FDTD. Here, we shall briefly review the approach in the weak-coupling limit, which serves as the foundation for the development of a theory for strongly coupled systems, such as plexcitons, as discussed later. In the framework of the continuum model, the polarizable continuum model (PCM) initially developed for molecules in solution has been extended to account for the electromagnetic response of the NP fully represented by the induced apparent charges defined on the NP surface.[145,146] The resulting PCM-NP is numerically implemented through surface discretization in small tesserae within BEM.[147,148] Each tessera is associated with a polarization charge that depends on the external potential due to the presence of an electromagnetic radiation or a molecular electron density. The PCM-NP equation in frequency domain reads

$$q(\omega) = -S^{-1}\left(2\pi\frac{(\varepsilon(\omega)+1)}{(\varepsilon(\omega)-1)}I + DA\right)^{-1}(2\pi I + DA)V(\omega) = Q(\omega)V(\omega) \quad (1)$$

where **S** and **D** are matrices that depend on the NP shape only, **A** is a diagonal matrix with elements equal to the tesserae areas and **V(ω)** is the external scalar potential oscillating at frequency ω. ε(ω) is the frequency-dependent dielectric function of the metal, which can be approximated and parametrized by a model function as the Drude-Lorentz one or associated to an empirical dielectric function. The right side of the equation, but for the external potential **V(ω)**, represents the BEM response matrix **Q(ω)**. This equation is valid in the quasi-static limit, that is the case when retardation effects are not included.

This is a frequency formulation that is most suitable for common spectroscopies such as UV-vis, IR and Raman. However, extension to an explicit time domain description is needed to extend the treatment to the newest experimental development in the field, such as ultrafast spectroscopies, also in the strong coupling regime, and investigation of charge injection events. The PCM-NP can be turned into time domain to treat the mutual interaction between a molecule and a NP during a real-time dynamic of the molecule wave function, by a Fourier transform of the PCM-NP equation[149–151]

$$q(t) = -\int_{-\infty}^{+\infty} Q(t-t')V(t')dt' \quad (2)$$

Practically, the time-dependent polarization charges are found to satisfy equations of motion (EOM) that are then solved numerically.[152]

To simulate the real-time coupled dynamics of a chemical moiety, described at the quantum level, and a metallic NP within the PCM-NP approach, an interaction term must be included in both the apparent surface charges EOM and the electronic time-dependent Schrödinger equation (TDSE). Within the latter, in the presence of a (time-dependent) external electromagnetic field, the electronic Hamiltonian can be expressed in very general terms as the sum of three contributions:

$$\hat{H} = \hat{H}_{mol} + \hat{H}_{mol-field} + \hat{H}_{mol-NP} \tag{3}$$

where $\hat{H}_{mol}$ is the electronic system Hamiltonian in vacuum, $\hat{H}_{mol-field}$ takes into account the interaction of the molecule with the external field, and $\hat{H}_{mol-NP}$ is the NP-molecule interaction. The time-dependent Schrodinger equation, governed by $\hat{H}$, can be easily propagated once an active space for the electronic dynamics is defined and the matrix elements of the operators present in the Hamiltonian are consequently computed.

Originally, this methodology was applied using an active space composed of the equilibrated molecular Hartree-Fock ground state and excited states obtained from a configuration interaction expansion limited to single excitations (CIS).[151,153] To overcome the limited accuracy of CIS description of excited states, and those of standard TD-DFT in the simultaneous description of localized and charge-transfer excitations, and of energy levels alignment, an active space derived within the MBPT formalism was proposed in Ref. [151] and applied to study the plasmon enhanced photocatalysis of $CO_2$ methanation on Rh nanocubes.[143] Reformulation of the PCM-NP for the strong coupling regime is described later in **section. 2.2.4.**

### 2.2.3 Accounting for atomistic features in the nanostructure

When the atomic features of the metal nanostructure play a crucial role in determining its plasmonic response – as in the presence of low coordinated sites, asymmetries, and atomic defects or in picocavities – atomistic approaches must be exploited. Among the atomistic approaches that have been developed, we here focus on the frequency-dependent fluctuating charges fluctuating dipoles (ωFQFμ) method.[154,155] This approach is specifically designed to include the physics underlying the plasmonic response of noble metal NPs: intraband transitions governed by Drude conduction mechanisms, and interband effects. In particular, the Drude model for intraband transitions is described by means of complex charges placed on each atom (q). They interact with complex dipoles also placed on each atom (μ), which account for an effective polarizability mimicking core electrons and interband transitions.[154] When an external electric field is applied to the sample, the atoms exchange charge via the Drude conduction mechanism, assisted by quantum tunnelling, which limits the charge transfer among the nearest neighbouring atoms with a typical exponential decay.[155] Charges and dipoles equations of motion are coupled to effectively account for electrodynamical interactions between the atoms, and are obtained by solving the following coupled sets of linear equations[156]:

$$-i\omega q_i(\omega) = \frac{2n_0\tau}{1-i\omega\tau}\sum_j\left[1-f\left(l_{ij}\right)\right]\frac{A_i}{l_{ij}}\left(\phi_j-\phi_i\right) \tag{4a}$$

$$\boldsymbol{\mu_i}(\omega) = \alpha_i^{\omega}\left(\boldsymbol{E}_i^{ext} + \boldsymbol{E}_i^{q} + \boldsymbol{E}_i^{\mu}\right) \tag{4b}$$

where $\tau$ is the scattering time, $A_i$ is the effective area of atom *i* through which the charge is flowing, and $n_0$ is the electron density. $\phi$ is the electrochemical potential acting on each atom, which accounts for interatomic interactions (charges and dipole moments) and their coupling to the external electric field. The function $f(l_{ij})$ ($l_{ij}$ is the distance between the *i*-th and *j*-th atoms) is a Fermi-like function introduced as a phenomenological description of quantum tunneling. $\boldsymbol{E}^{\mu}$ and $\boldsymbol{E}^{q}$ are the electric fields generated by the dipoles and by charges acting on atom *i*, respectively. Finally, $\alpha_i^{\omega}$ is the frequency-dependent atomic polarizability which models interband transitions. Note that the parameters entering ωFQFμ have a clear microscopic physical meaning, and their numerical values can be obtained from experiments or fitted to reproduce higher-level results (see Ref. [154]). The model can remarkably capture plasmonic properties of systems featuring sub-nanometre junctions,[154,155] defects,[157,158] bimetallic compositions,[159] and solvent effects.[160] Furthermore, by an EOM formulation, the method has been extended to the real-time domain,[156] allowing for an accurate simulation of the time evolution of the plasmonic response on the ultrafast timescale. The coupling of the method with a quantum mechanical (QM) treatment of a nearby molecular system allows for the simulation of surface-enhanced spectroscopies, such as Surface-Enhanced Raman Scattering (SERS)[161] and SE Fluorescence (SEF).[162] Extension of this formulation to the strong coupling regime is an active line of research.

### 2.2.4 Modelling quantum plasmons

A quantum-mechanical framework for the nanostructure is essential for investigating plasmon-exciton hybrids under strong coupling conditions. The ground and excited states of the plasmonic material system are combined with those of the molecular (excitonic) partners, thereby defining the Hilbert space spanned by the resulting plexcitonic wavefunctions. While the transition energies and properties of organic chromophores can be routinely computed using various levels of electronic structure theory - most commonly TDDFT - analogous information for the plasmonic resonances of metal nanostructures remains less readily accessible. A full *ab initio* treatment of the plasmonic subsystem is in most cases unfeasible due to the typical size of nanostructures.

Consequently, deriving a plasmonic wavefunction based on a continuum description becomes a crucial step in modelling plexcitonic systems. In particular, a quantized representation of plasmon modes has been proposed and validated within the framework of PCM-NP, leading to the Q-PCM-NP approach.[163] More precisely, the BEM response matrix in the PCM-NP approach[149] is the starting point for deriving an expression for polarization charges in a form that bears a close analogy to a sum-over-states quantum linear response function, involving discrete plasmonic states. By considering the Drude-Lorentz model dielectric function, the following expression is obtained for the diagonalized PCM-NP response function,[163]

$$\boldsymbol{Q}_{kj}(\omega) = -\sum_p \left( \frac{\langle 0|\hat{q}_k|p\rangle\langle p|\hat{q}_j|0\rangle}{\omega_p+\omega+i\frac{\gamma_p}{2}} + \frac{\langle p|\hat{q}_k|0\rangle\langle 0|\hat{q}_j|p\rangle}{\omega_p-\omega-i\frac{\gamma_p}{2}} \right) \quad (5)$$

where the index $p$ identifies different plasmon modes, while $k$ and $j$ run over surface tesserae. As usual, the denominator accounts for the excitation energy of each plasmon resonance, while the transition matrix elements in the numerator provide a set of surface charges describing the full transition density associated with the excitation of the corresponding plasmon mode. These transition charges can be employed to evaluate the coupling with molecular transitions, either directly or approximately via the corresponding transition dipole moment.[31,164] Recently the quantized description of plasmonic nanostructures has been extended to incorporate a generic form of the dielectric function.[165] Additionally, a quantized formulation based on atomistic approaches to plasmonic NPs is also being pursued.

# 3 Applications: Theory and Experiment

## 3.1 Simulating molecules coupled to plasmonic nanostructures

### 3.1.1 Laying the Groundwork: Weak-Coupling Calculations Toward the Strong-Coupling Regime

The architectures that are known to potentially provide strong coupling between molecular electronic excitations and plasmonic resonances (nanocavities formed with a tip on a substrate, plasmonic NP arrays and aggregates, nanoparticle-on-mirror geometries) can yield a plethora of intriguing phenomena in the weak coupling regime as well. They compete with strong coupling effects when the system is on the onset of strong coupling regimes. As such, they should also be simulated to have a complete understanding of the behaviour of such systems. For example, it has been experimentally demonstrated that tip-on-substrate nanostructures, capable of confining the electromagnetic field down to the nanoscale, enable photoluminescence with sub-molecular spatial resolution.[166] From a theoretical standpoint, tip-enhanced photoluminescence can be described in terms of the modified quantum yield and the enhanced molecular absorption.[167,168] Combining the PCM-NP determination of these quantities, the tip-enhanced photoluminescence of zinc-phthalocyanine set between a silver substrate and a silver nanotip could be simulated.[166,169] A similar approach has been applied to the computation of the electronic energy transfer rate between two phthalocyanines complexes mediated by a silver nanotip.[170] The calculations provided an explanation for the apparent exponential decay of the excitation energy transfer efficiency measured experimentally, that would be generally taken as a signature of Dexter-like excitation energy transfer mechanisms, but instead can be rationalized in terms of the electromagnetic interplay between plasmons and molecular excitations. The capability of the tip-on-substrate (that effectively realizes a plasmonic nanocavity) to qualitatively modify the energy transfer rate between a donor and an acceptor is reminiscent of the capability of larger cavities to modify energy transfer on a long distance, see sect. **3.2.1**. Many other properties can also be calculated by a recent time-domain formulation of PCM-NP,[151,171] such as molecular absorbance, Raman scattering, and electronic circular dichroism (ECD) spectra mediated by plasmonic systems.[172,173]

It should not be forgotten that local electromagnetic amplification and strong coupling are not the only phenomena that may take place when molecules are located close to NPs or in nano-/pico-cavities. Within tens of femtoseconds after excitation, a plasmonic resonance can decay into highly energetic, non-thermalized hot electrons and holes. These out-of-equilibrium carriers and hybrid states can be exploited in a wide range of applications such as photocatalysis, solar energy conversion, and molecular sensing.[174,175] Multiscale approaches have proved effective in studying injection of such carriers in molecules for plasmon-driven photocatalysis. Quantum-continuum models have been applied to Rh nanocubes to study $CO_2$ reduction to methane, revealing that plasmon-induced hybrid states facilitate hole injection from the metal to the reaction intermediate, selectively weakening C–O bonds while preserving C–H bonds.[143,176] This explains experimentally observed enhancements in selective $CH_4$ production.[177] Similar techniques showed also that AuRh alloys and selected Cu-based systems featuring specific coordination sites are promising candidates for plasmon-mediated $CO_2$ reduction to methane.[178] While the experimental regimes of strong coupling and hot-carrier injections are not necessarily overlapping, allowing focused investigation of one regime or the other, recent proposals[179] of using the strong coupling to control hot-carrier injection opens new lines of research for multiscale simulations.

### 3.1.2 When atomistic description is essential: the peculiar case of picocavities

To illustrate the potential of fully atomistic modelling, we consider the case of plasmonic picocavities—optical cavities where single-atom defects (adatoms) confine light down to picometer scales with effective mode volumes below 1 nm$^3$. In such systems, strong light-matter coupling can therefore be realized. To gain insight into the morphology of the electric field induced in plasmonic picocavities, it is necessary to resort to fully atomistic modelling approaches, such as ωFQFμ. These methods can treat systems of very large size, a fundamental requirement to avoid artificial boundary effects that cannot be reliably handled at the ab initio level. In Ref. [158], we investigated the morphology of the electric field in picocavities constructed by introducing a single adatom in the first outer atomic shell of one of the two metallic plates forming the nanocavity. The resulting structure comprises approximately 174,000 atoms and closely resembles the NPoM geometry, which is the prototypical configuration widely exploited in experimental studies. **Figure 5** compares the optical response of nanocavities (top) and picocavities (bottom). The absorption spectra are substantially identical (**Fig. 5a**), both displaying a main plasmon resonance peak at about 2.16 eV, because a single-atom defect does not alter the global plasmonic response of such large structures. The induced field enhancement in the gap, $|E|/|E_0|$ shows a maximum around 2.12 eV for both cavities (**Fig. 5a**, blue line), with relatively modest values (<10) due to the limited cavity thickness and absence of sharp features. Yet, while the global spectra are unaffected, differences emerge in the field morphology, which is associated with the induced charge density. The induced charge density maps (**Fig. 5b and c**) reveal that the nanocavity supports the boundary dipolar plasmon (BDP) mode, with homogeneous charge distribution across the plates. In contrast, the adatom in the picocavity generates a pronounced and asymmetric local dipole (**Fig. 5c**, bottom). The consequences of such an inhomogeneous charge distribution are evident in the electric field maps (**Fig. 5d**). The nanocavity exhibits a nearly uniform enhancement in the gap, resembling the field of a parallel-plate capacitor. In the picocavity, the field is uniform away from the defect. Still, it becomes strongly inhomogeneous in its proximity, showing cylindrical symmetry with enhanced fields near the adatom's van der Waals radius and depletion directly above it. Such a morphology, which can be identified only by exploiting ωFQFμ, directly impacts light-matter interactions at the nanoscale, with implications for molecular plasmonics and other applications such as SERS.

### 3.1.3 Modelling plexciton properties and dynamics

Theoretical modelling of plexcitonic nanosystems primarily focuses on the resulting hybrid states to gain insights into the distinctive photophysical and photochemical behaviours observed experimentally. Many experimental systems discussed in this work involve large ensembles of chromophores coupled to plasmonic platforms, where strong coupling emerges as a collective phenomenon. This complexity renders full *ab initio* calculations of the entire system computationally infeasible. To date, various formulations of ab initio quantum electrodynamics methods have been successfully applied only to systems comprising one or a few small molecules.[180–183] Only recently have generalizations to collective effects started to appear.[184]

Consequently, the hybrid states of interest are most commonly explored through the construction and diagonalization of effective model Hamiltonians, such as the Tavis-Cummings model,[185–187] which often entails a simplified description of the interacting components. In particular, the molecular subsystem is typically represented as a structureless entity characterized by a collective transition dipole, thereby implicitly neglecting molecular-level details.

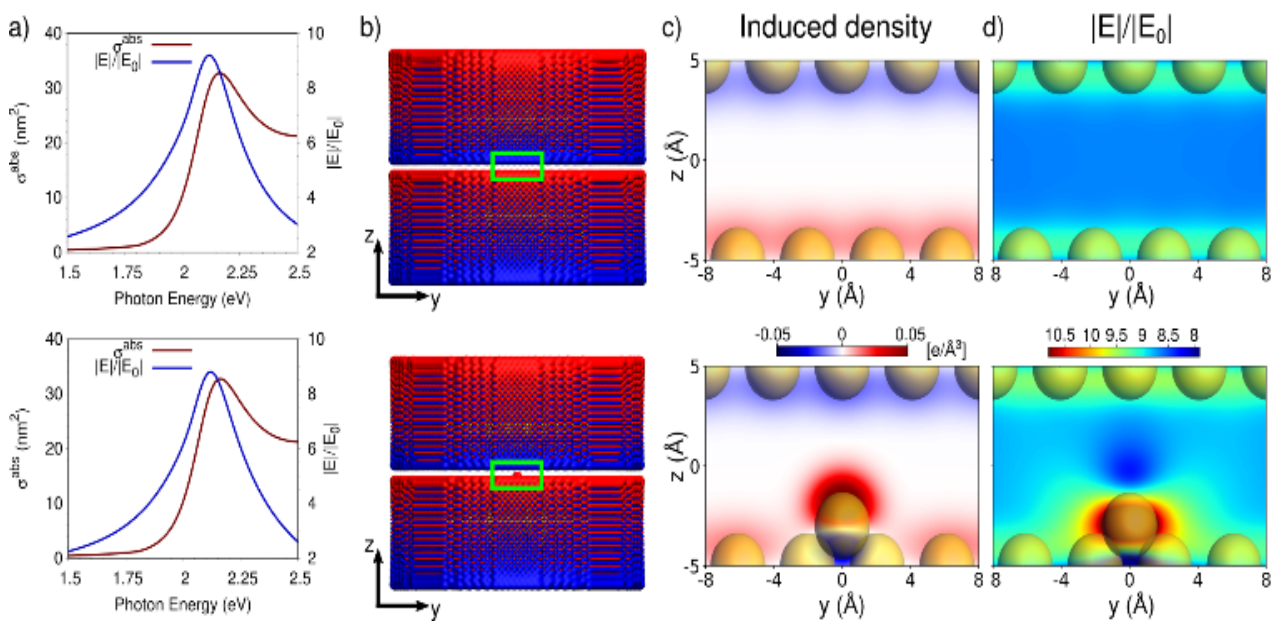

**Figure 5**. (a) ωFQFμ absorption cross sections and field-enhancement factors, (b) induced charge density calculated at the plasmon resonance frequency, and zoom maps of the region highlighted in panel b of (c) induced density and (d) electric field enhancements for nano (top) and picocavity (bottom). Reproduced with permission from ref.[158] Copyright 2025 Author(s), licensed under a CC-BY Creative Commons Attribution 4.0 License.

Recent advances in these simplified models aim to incorporate greater physical detail. For instance, the individuality of each molecule-plasmon coupling - which critically depends on the spatial position and orientation of the molecule relative to the plasmonic platform - has been taken into account in studies of plexcitonic hybrids formed by non-aggregated porphyrins non-covalently attached to colloidal gold nanospheres.[24,164] The construction of effective plexcitonic models has incorporated experimental parameters and classical molecular dynamics results, providing insight into the precise structural arrangement of the system.[188] Moreover, a new strategy for calculating individual molecule-plasmon electrostatic coupling has been developed, going beyond the point-dipole approximation and potentially accounting for the non-uniform shape of molecular transition densities.[164]

These theoretical improvements have been pursued not merely for increased realism, but to extract deeper insights and elucidate specific experimental observations. In particular, they enabled the identification of local inhomogeneities in the molecular contributions to the resulting plexcitons, highlighting the role of (nano)cavities in complex NP assemblies (see **Figure 6**). Analogous results have been obtained for spiropyran photo-switches located near the long-axis tips of aluminium ellipsoids.[31] These findings are consistent with the electric field profile of electromagnetic hotspots and have significant implications for energy flow, light harvesting, and photochemical dynamics. Depending on the molecular contributions to the plexcitonic wavefunctions, the energy provided by hybrid-state excitation is more likely to be funnelled toward privileged sites rather than evenly distributed across the ensemble of chromophores. These sites may thus favour the conversion of excess energy into chemical transformations.

This picture reconciles the collective and local character[189] of plexcitonic systems: the need to include a large number of molecules to access the strong coupling regime as a collective effect does not contradict the existence of local inhomogeneities and non-uniform molecular contributions. Additionally, the influence of intermolecular interactions on the structure of the so-called "dark state" reservoir, as well as the ability of plasmons to mediate coupling between excitations of different character, has been revealed - providing theoretical support for experimental findings on ultrafast energy transfer in the same system.[25,164] These observations stem from explicitly accounting for the multilevel structure of many organic dyes, which often possess several electronic excited states within a narrow energy range, and for intermolecular (excitonic) interactions even in loosely organized molecular assemblies.[164]

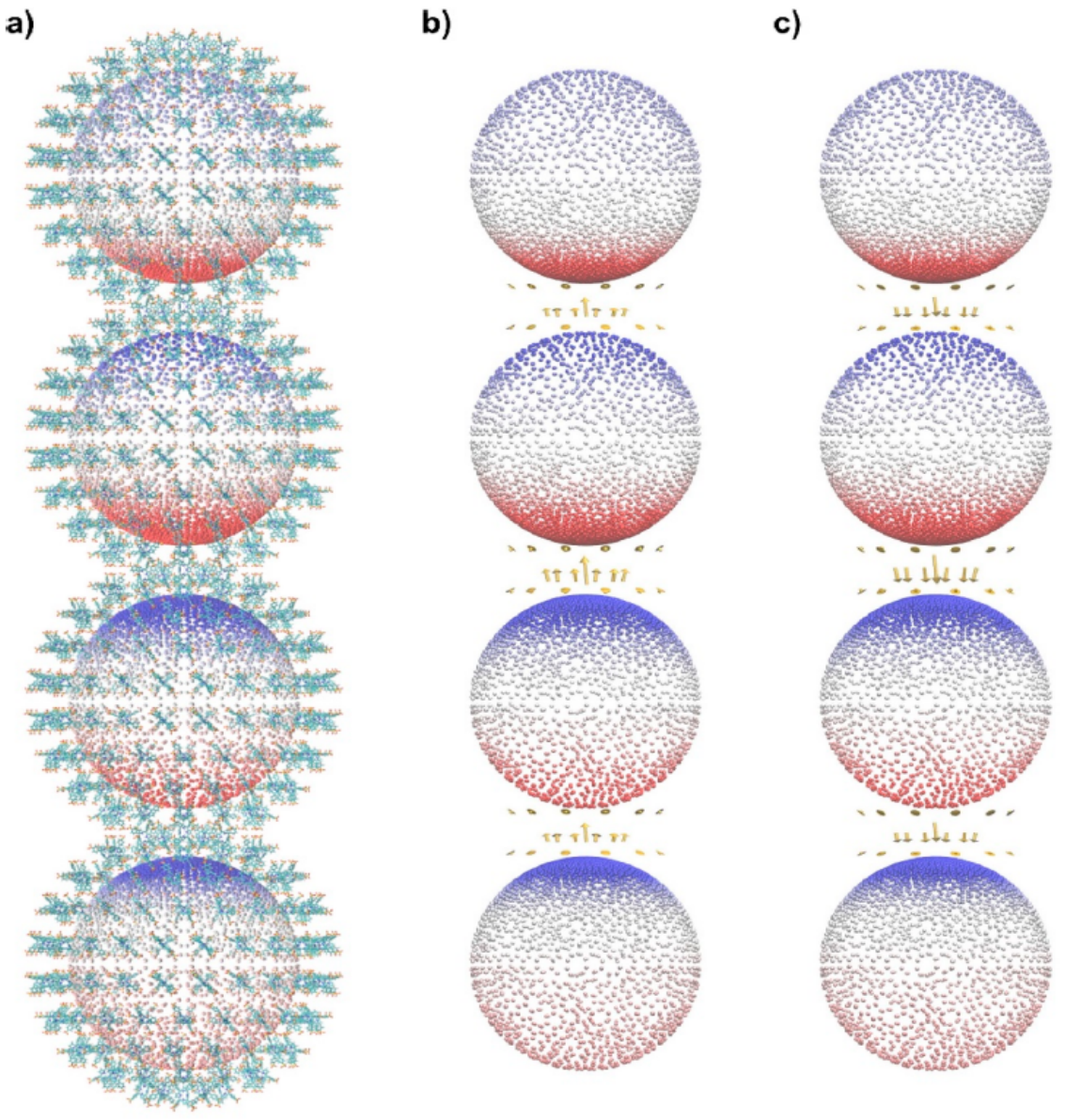

**Figure 6**. (a) Arrangement of a molecular layer around an aggregate of four gold nanoparticles used to simulate plexciton formation with the quantized version of PCM-NP, Q-PCM-NP. (b) Representation of the molecular contributions to the wavefunction of the lowest plexciton. The molecules in the cavities between the nanoparticles contribute the most. (c) The same for the upper plexciton. Reproduced with permission from ref. [164] Copyright 2024 American Chemical Society.

These hybrid modelling approaches are also pivotal to clarifying the ultrafast dynamics after the excitation of the plexcitons. For example, in ref.[31] they were used to rationalize the results of pump-probe experiments on a molecular photo-switch (spiropyran) strongly coupled to an aluminium nanoantenna (see **Figure 7**). The experimental data could be reproduced nicely once one assumed that the plexciton excited by the pump collapsed within a few tens of fs into a molecular excitation that in turn evolved following a nuclear relaxation pathway in the electronic excited state. Notably, even if the initial hybrid state is rapidly lost, plexcitonic effects are still relevant both to modify the decay time of the molecular excited state and in the measurement itself. The probe pulse, in fact, creates a new plexcitonic state in which the excited molecule cannot fully participate because the molecular Stokes shift is developing in time. Hybrid models were also fundamental to compare a fully quantum vs a semiclassical description of the light excitation of the electronic state of a chromophore close to a metal NP.[190]

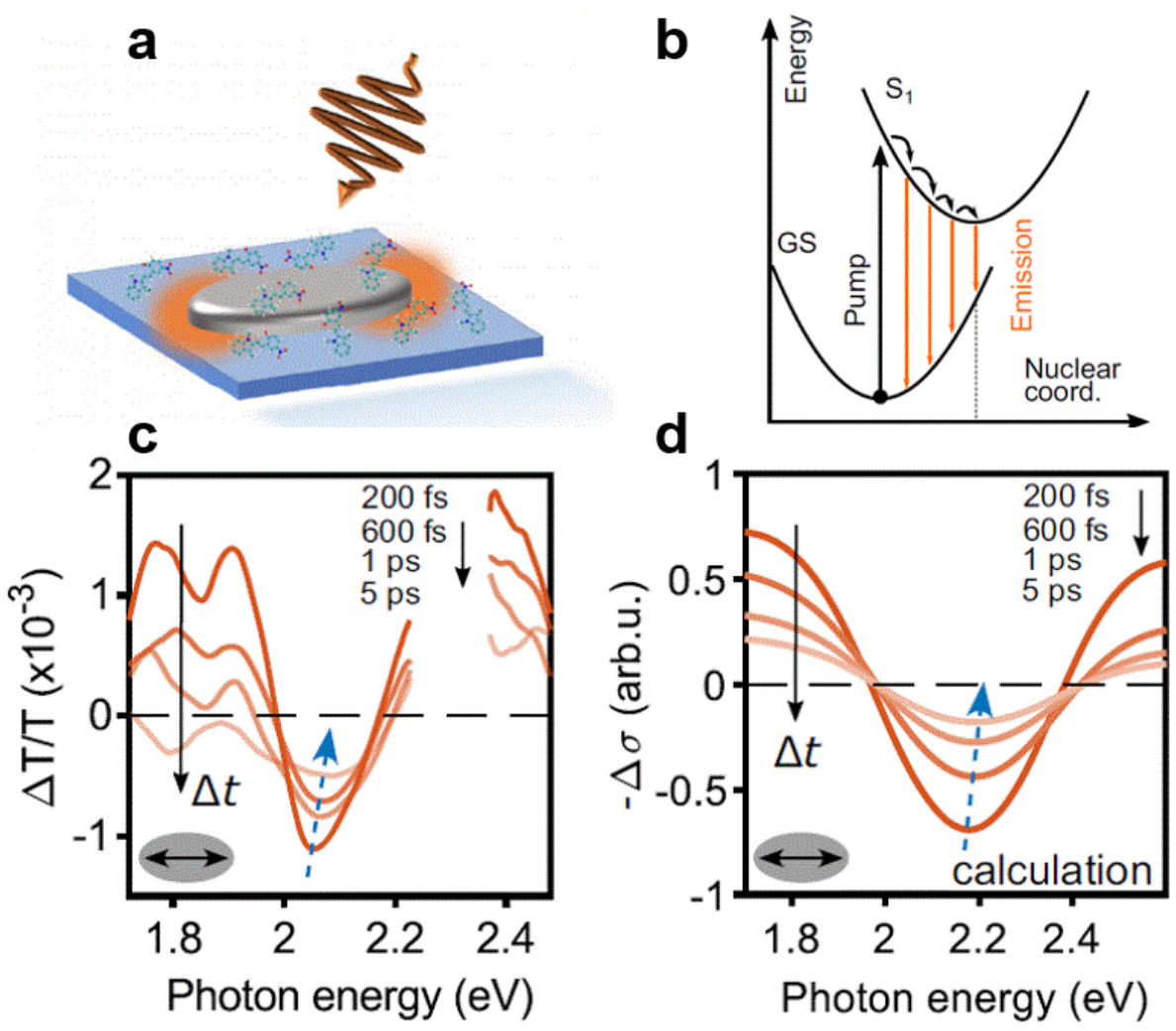

**Figure 7**. (a) Experimental and theoretical investigation of a plexcitonic system obtained by covering an array of anisotropic aluminum nanoellipse antennas with a matrix containing spiropyran molecules. The pump-probe transient absorption measurements could be reproduced by assuming the collapse of the collective plexciton on a single molecule tens of fs after its formation; (b) the molecule then relaxes within its electronic excited state changing the frequency of the stimulated emission and excited state absorption signal. Taking into account ground state bleaching, this shifted stimulated emission and the excited state absorption to the double-excitation plexcitonic manifold, the experimental time-resolved transient absorption measurements (c) could be reproduced by the simulation results based on the Q-PCM-NP model (d). Adapted with permission from ref. [31] Copyright 2023 Author(s), licensed under a CC-BY Creative Commons Attribution 4.0 License.

## 3.2 Experiments on polariton-mediated optoelectronic properties of organic molecules and 2D materials

### 3.2.1 Polariton-mediated long-range ultrafast energy delocalization in a strongly coupled microcavity

The hybridized nature of polaritons makes them particularly attractive compared to uncoupled states, especially when considering energy exchange pathways in organic systems. The delocalized character of polaritons, inherited from their photonic component, enables coherent, long-range interactions between molecular species well beyond the energy transfer range limit imposed by the Förster radius, with major implications for polariton-based devices. By extending the transfer length between localized excitonic states, polaritons can overcome the performance limitations of conventional organic semiconductors, where exciton diffusion lengths are typically restricted to just a few nanometres. This opens the way to light-harvesting and opto-electronic devices with functionalities otherwise inaccessible. Within this framework, a recent study[5] (**Figure 8**) tracked in real time the mechanism of long-range energy transfer mediated by polaritons between donor and acceptor molecules placed inside an optical cavity and separated by a mesoscopic distance of 2 μm. Using 2DES, the authors revealed the ultrafast energy flow dynamics and showed quasi-instantaneous energy delocalization across polariton states, which acts as a direct bridge between molecular excitons located in distinct layers.

The investigated system consisted of a metallic (Ag) Fabry-Pérot microcavity containing two thin films of J-aggregated organic semiconductors (TDBC and NK2707) separated by a 2-μm inert, transparent spacer (polystyrene), whose mesoscopic thickness, much larger than the Förster radius, prevents direct energy transfer between the two dyes (**Fig. 8a**). Inside the cavity, three polariton branches (upper, UPB, middle, MPB, and lower, LPB) were identified, arising from the hybridization of a cavity mode with the donor, with both donor and acceptor, and with the acceptor, respectively (**Fig. 8b**). Owing to the 15-fs temporal resolution of the 2DES measurements, coherent coupling among the three polariton branches could be directly resolved, while retaining full spectral information (**Fig. 8c**). When the uncoupled (either donor or acceptor) exciton reservoirs were photoexcited, cross-peaks emerged at the probe energies of the middle and either the lower or upper polariton, consistent with the dominant matter contribution to the hybrid state. In contrast, pumping the upper polariton produced cross-peaks simultaneously at the middle and lower branches, with an almost instantaneous, sub-20 fs rise time (**Fig. 8d**). Notably, no signal from the lower polariton was observed following excitation of the high-energy exciton reservoirs. The quasi-instantaneous (within the limit of the experiment resolution) rise-time of the cross-peaks under upper polariton photoexcitation was attributed to coherent mixing of polariton states and rapid energy delocalization across the entire cavity. This coupling mechanism mediated energy transfer between the molecular species even though they are physically separated by 2 μm, far exceeding Förster-limited distances. Control measurements on the same multilayer film (i.e., J-aggregated TBDC and NK2707 dye layers, separated by the spacer film, **Fig. 8a**) outside the cavity showed no fingerprint of energy transfer, confirming unambiguously the pivotal role of polariton-mediated coupling. The experimental findings were supported by a time-resolved, fully coherent broadband model of polariton interactions. While theoretical calculations were unable to provide a quantitative prediction, they reproduced correctly the ultrafast formation and spectral line shapes observed in the 2DES maps, validating the interpretation of instantaneous energy delocalization between the different exciton components driven by a single cavity mode.

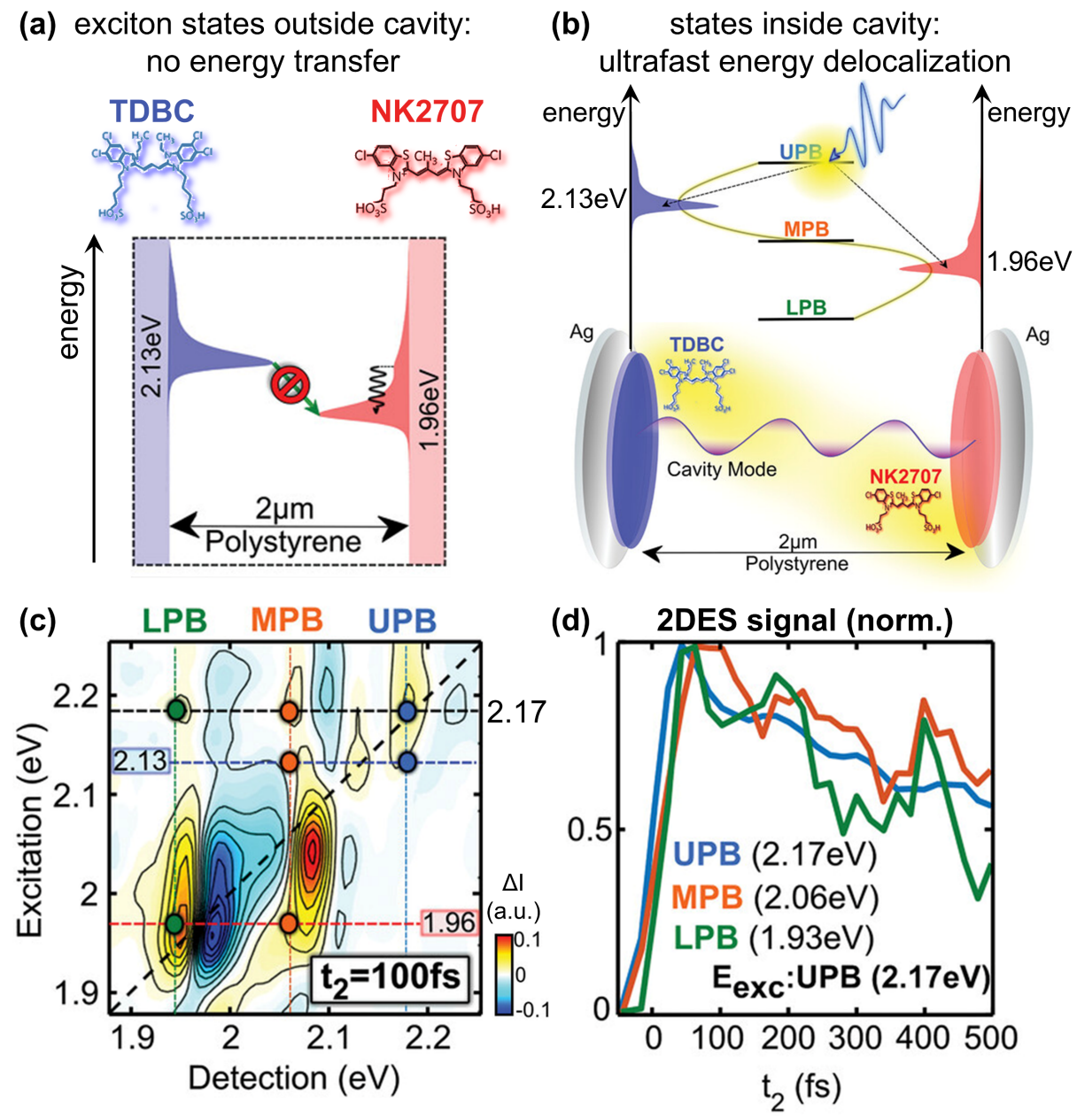

**Figure 8**. Ultrafast energy delocalization in a strongly coupled microcavity. (a) Schematic of the considered multilayer film outside the cavity, along with the chemical structure and steady-state absorption of the J-aggregated dyes. (b) Illustration of the energy level scheme of the organic microcavity in the strong coupling regime, where the energy path providing direct connection between donor and acceptor molecules is highlighted. (c) Purely absorptive 2DES map at a representative waiting time ($t_2$=100 fs). (d) Temporal traces extracted by exciting the UPB (2.17 eV) and detecting the UPB (blue), MPB (orange), LPB (green) states. Adapted with permission from, ref. [5] Copyright 2024 The Author(s), licensed under a CC-BY Creative Commons Attribution 4.0 License.

### 3.2.2 Polariton-mediated conductivity in organic mirrorless cavities

In addition to polariton-mediated energy transfer discussed above, the delocalized nature of polaritons can enhance charge transport in optoelectronic devices.[79] The effect of strong coupling on electronic conductivity was first investigated by coupling n-type organic semiconductors to a plasmonic hole array,[33] followed by investigations of photoconductivity in p-type organic semiconductors[76] and in an inorganic two-dimensional semiconductor ($WS_2$).[80] The question of whether such effects can be observed in planar, open cavities is important as mirrorless configurations may offer a simpler way of boosting functional material properties.

Kaur et al. investigated the possibility of conductivity enhancement under strong coupling by coupling the organic semiconductor, perylene diimide (PDI), to a mirrorless cavity.[78] The experiments were conducted in a bottom-gate, bottom-contact metal oxide semiconductor field effect transistor (MOSFET) configuration (**Figure 9a**). In this structure, the PDI dye (active layer) is spin coated on a silicon/silicon dioxide surface which allows FET-mode operation. Notably, it was found that light confinement due to the refractive index distribution in the multilayer ($Si/SiO_2$/PDI/air) modifies electronic conduction in the active layer. Here, silicon acts as the bottom

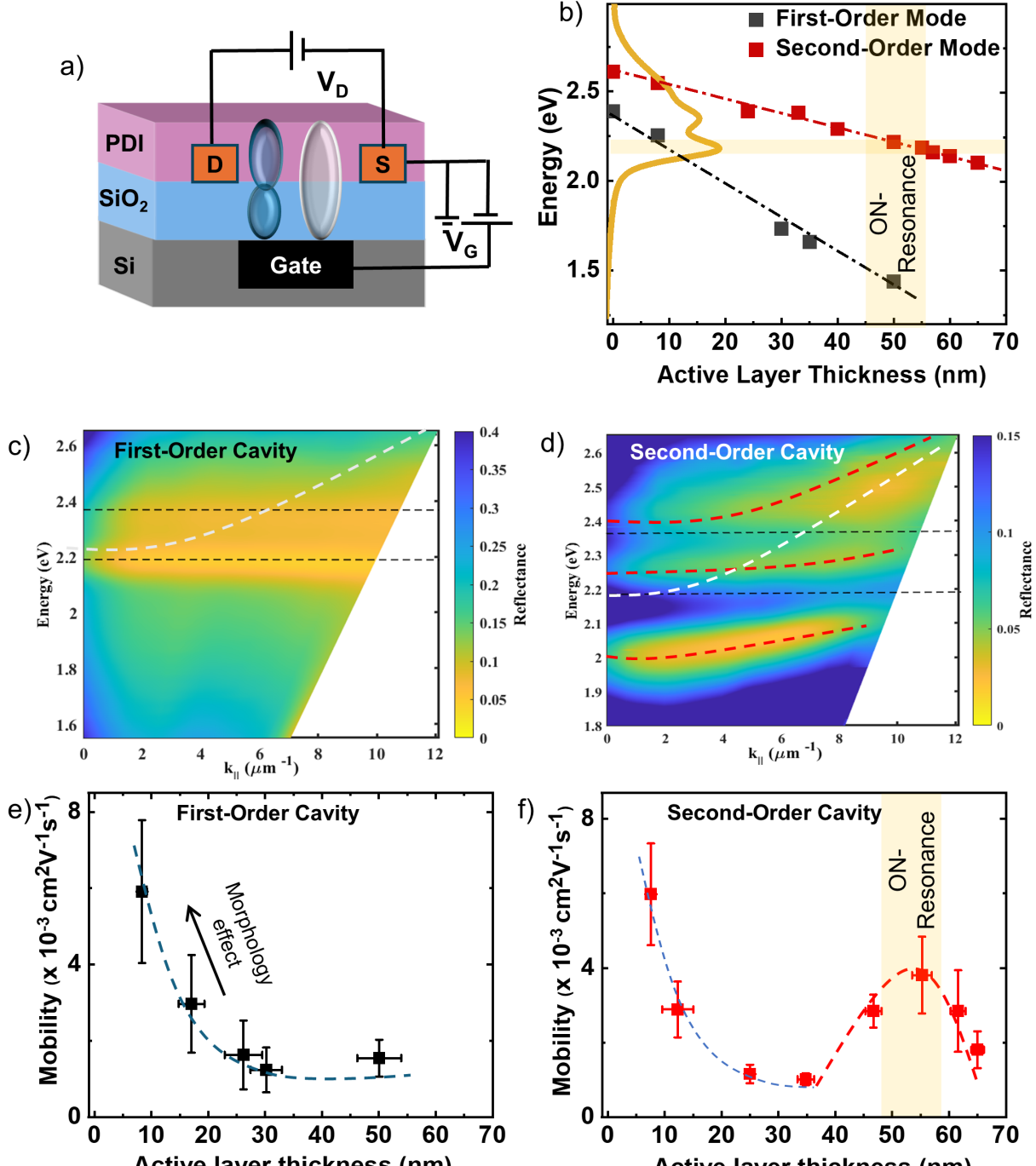

**Figure 9**. Enhanced charge transport under electronic strong coupling. (a) Schematic of MOSFET device in a mirrorless cavity configuration. The active layer (PDI) is represented in pink. First-order and second-order leaky cavity modes are formed depending on the cavity thickness. (b) Cavity mode energy versus active layer thickness measured experimentally. The dotted/dashed lines are guide to the eye. (c,d) Fourier-plane dispersion (reflectance) with (c) no polaritonic branching for first-order mode and (d) branching observed for the second-order mode. The dispersion simulated by transfer matrix method is represented by red lines. In (c,d), dashed lines indicate the absorption maxima of PDI (black color), and the dispersion of the empty cavity mode (white color). (e,f) Electron mobility for (e) first order and (f) second order cavity MOSFET devices. An enhanced mobility is measured under ON-resonance condition for the second-order cavity. Dotted/dashed lines are guide to the eye. Adapted with permission from ref.[78] Copyright 2023 Wiley-VCH GmbH.

'mirror' and the relatively high refractive index of PDI (1.7) allows leaky optical modes with low to moderate quality factors to be formed. Two cavities were studied to identify their effect on PDI conductivity, one in which PDI is coupled to the first order cavity mode and the other where PDI is coupled to the second order mode. **Figure 9b** summarizes the active layer thickness for which the cavities become resonant with the PDI exciton (2.2 eV and 2.35 eV) at normal incidence. The first order mode, which has larger losses, couples weakly to the PDI exciton absorption, whereas the second order mode offers better coupling strength. Reflectance measurements using Fourier microscopy were able to accurately measure the dispersion for first and second order cavity coupled devices from the same microscopic region where conduction is measured between gold electrodes (**Figure 9c** and **Figure 9d,** respectively). Interestingly, the second order mode (with an improved cavity quality factor) couples more strongly with the PDI excitons and the anti-crossing of dispersive polaritonic states was observed via Fourier microscopy (**Figure 9d**). Transfer matrix method simulations[191] further verified the Rabi splitting in the hybrid devices coupled to the second order mode. Electron mobility measurements in a set of devices prepared by varying the thickness of PDI demonstrated a two-fold increase when the second order cavity mode becomes degenerate with the exciton absorption (**Figure 9f**). This mobility enhancement effect was not observed for the first order cavity, where anti-crossing of dispersions was absent (**Figure 9e**). Overall, the experiments suggested that strong coupling of the excitonic transition modifies electron transport without physically modifying the conducting layer. The outcomes lend support to the idea of vacuum engineering of materials, i.e., the modification of material properties by simply coupling to vacuum electromagnetic modes.[1,7,192]

### 3.2.3 Identifying the dark-strong coupling regime in cavity-coupled monolayer $WS_2$

The conventionally adopted strong coupling condition requires that the rate of energy exchange between interacting light and matter states is greater than their dissipation rates.[3] This condition ensures that Rabi oscillations can take place before the excitation energy decays from the system. However, the condition for the existence of polaritonic states is different and is given as

$$g > \frac{|\gamma_c - \gamma_x|}{4} \tag{6}$$

where $g$ is the coupling strength and $\gamma_c$ ($\gamma_x$) represent the dissipation rates of the cavity (excitonic) states.[3,193] This condition relates to the emergence of non-degenerate eigenenergy levels of the coupled system and hence also represents the weak to strong coupling transition that occurs at an exceptional point (EP).[194] At the EP, the coupling strength becomes $g_{EP} = |\gamma_c - \gamma_x|/4$. Therefore, $g > g_{EP}$ ($g < g_{EP}$) gives the condition for strong (weak) coupling in this picture.

It is notable that the condition in equation 6 is easier to satisfy than the usual strong coupling condition, which can be written as $g > (\gamma_c + \gamma_x)/4$. In other words, polaritonic states can exist even though the Rabi splitting may not be larger than the losses. This is referred to as intermediate coupling in recent literature, especially in plexcitonic systems.[195,196] As an extreme case of the intermediate coupling regime, it was recently predicted that polaritonic states can exist even in the complete absence of a spectrally measured Rabi splitting (SRS), in the so-called dark-strong coupling regime.[100] Due to absence of an observable splitting in the optical spectrum, identifying this coupling regime and differentiating it from weak coupling becomes challenging. Johns et al. investigated such as system where a monolayer of $WS_2$ was placed in a planar, open cavity formed by a layer of hBN on a silver mirror (**Figure 10a**).[4] An analytical approach based on the coupled oscillator model was employed to predict the existence of an EP for this system at an hBN thickness of $L \approx 1100$ nm (**Figure 10b**). The location of the EP (in other words, the onset of dark-strong coupling here) is determined by the terms $g$ and $\gamma_c$ in equation 6, both of which depend on the cavity thickness $L$. Furthermore, by evaluating the conditions for SRS to appear in the excitonic and cavity channels of the absorption spectrum, it was verified that when $g \gtrsim g_{EP}$ ($L \gtrsim$ 1100 nm), the system lies in the dark-strong coupling regime,[100] where non-degenerate polaritonic levels exist but are nonetheless hidden in the optical spectra by losses. To experimentally probe the system, Fourier imaging was employed to measure the reflection and PL dispersion of the exciton-cavity system for a range of hBN thicknesses on either side of the EP ($L$ = 475 nm and $L$ = 1720 nm, see **Figure 10c,d**). As theoretically predicted, an SRS was not observed in the system from either absorption or PL measurements. However, a strongly dispersive emission from multiple branches was observed for samples with $L$ > 1100 nm, pointing to a possible polaritonic origin of the emission (**Figure 10d**). Comparison with the dispersion calculated by a quasi-normal mode (QNM) analysis showed excellent agreement with the measured dispersion of the PL signal. Notably, the QNM analysis showed the presence of a splitting in the system only for $L$ > 1100 nm (**Figure 10e,f**). The maximum value of the splitting is estimated to be 17 meV, which is well below what can be detected in the optical spectrum due to the linewidths of the cavity and exciton (≈ 28 meV). Nonetheless, the presence of polaritonic states could be inferred experimentally by the dispersive emission from the samples. These results suggest that polaritonic behaviour is possible even when a coupled system does not reach the usual strong coupling condition. This may be especially relevant in open cavity systems with relatively low coupling strengths, which could support polaritonic states near an EP. Moreover, the study demonstrates that resolving PL signals in momentum space via Fourier imaging can provide valuable insight on the nature of polaritonic states.

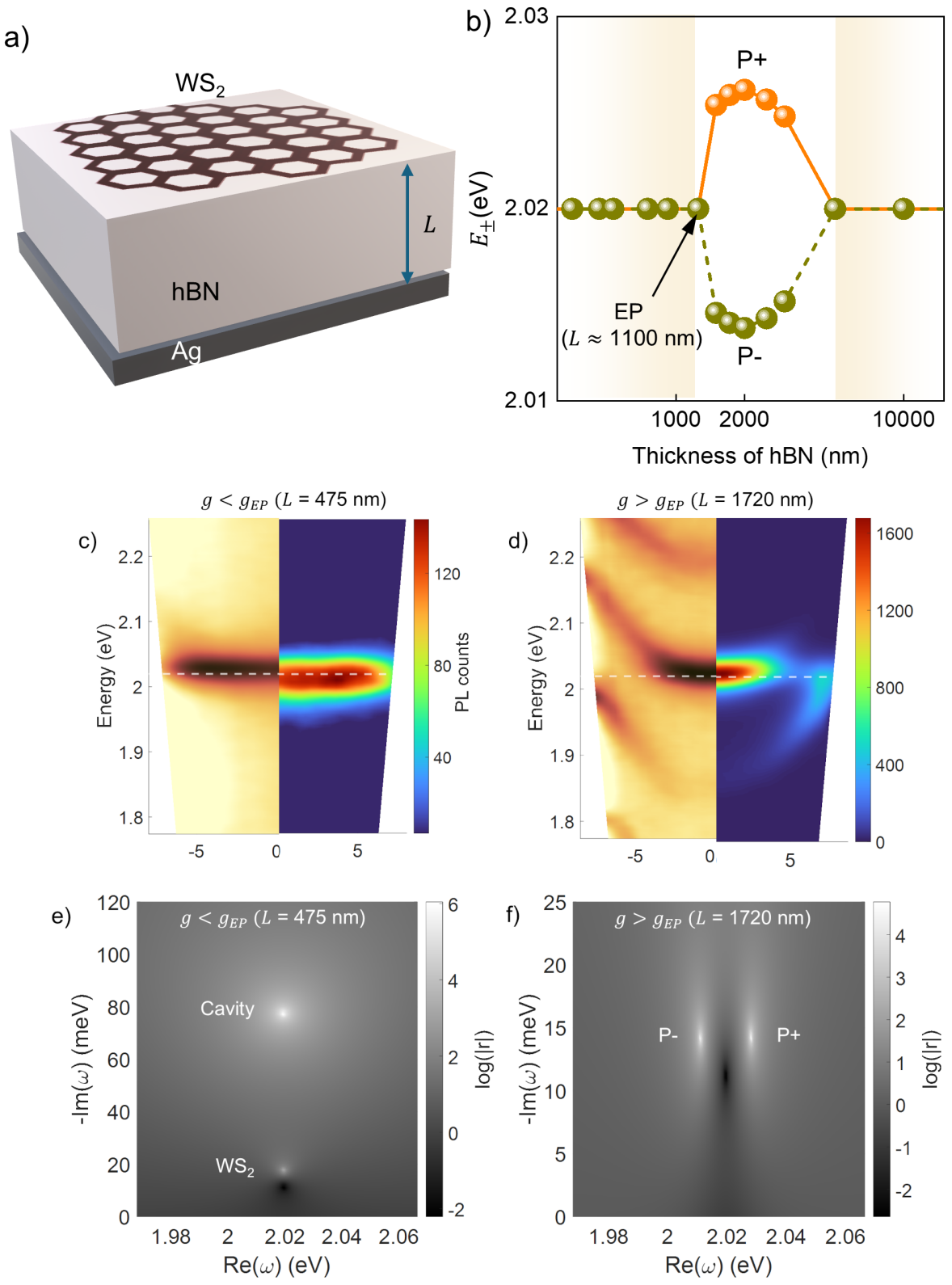

**Figure 10**. Identifying polaritonic states in the dark-strong coupling regime. (a) Schematic of the exciton-cavity system with $WS_2$ on top of an hBN layer on a silver substrate (b) Calculated eigenenergy values of the exciton-cavity system as a function of hBN thickness. The presence of two exceptional points bounding a region of non-zero level splitting is observed. (c,d) Reflectance (left) and PL (right) for hBN thickness ($L$) of (c) 475 nm ($g < g_{EP}$) and (d) 1720 nm ($g > g_{EP}$) showing contrasting behaviour in PL dispersion (see text). (e,f) Results of quasi-normal mode analysis for $L$ = 475 nm (e) and $L$ = 1720 nm. The complex frequency maps from quasi-normal analysis show bright regions indicating poles, which show no splitting for L = 475 nm and identifying a level splitting when $L$ = 1720 nm. Adapted with permission from ref.[4] Copyright 2025 Wiley-VCH GmbH.

## 4. Conclusions and Outlook

Despite significant progress in understanding hybrid light-matter interactions, several fundamental challenges remain before the full potential of polaritonic and plexcitonic systems can be unfolded. Addressing these challenges requires tightly integrated advances in both experiment and theory, as many open questions sit at the interface between material complexity, electromagnetic confinement, and ultrafast quantum dynamics. Here, we first list a few existing challenges and outline possible routes to overcome them. Following this, we discuss future perspectives and emerging research directions that we believe may help address them.

**1. Vibronic structure and coherence in strongly coupled systems.** A key open question concerns how molecular vibrations influence coherence, transport, and chemical reactivity under strong coupling. Experimentally, accessing this regime demands ultrafast techniques with high spectral resolution capable of disentangling vibronic contributions from polariton dynamics. Theoretically, fully capturing vibronic-polaritonic interactions requires models that combine an accurate description of the vibro-electronic structure with a realistic representation of the molecular environment and cavity field. Progress will hinge on joint experimental-theoretical frameworks capable of validating vibronic signatures across multiple platforms, ultimately enabling rational design of systems where vibrations are used to stabilize coherence, guide population flow, or tailor chemical pathways.

**2. Ultrafast nonlinear spectroscopy and its theoretical interpretation.** As multidimensional spectroscopies become central to the field, interpreting their signals in strongly coupled systems remains a major bottleneck. The challenge is twofold: experiments must resolve fast dephasing, overlapping resonances, and dark-state dynamics, while theory must model nonlinear optical responses in systems where thousands of molecules interact with structured photonic reservoirs. Development of multiscale quantum/classical frameworks that can simulate transient and 2D spectra will provide the missing link, allowing direct comparison between simulated and measured signals and enabling extraction of the microscopic mechanisms behind coherent and incoherent polariton dynamics.

**3. Temporal resolution and field-resolved photophysics.** Emerging field-resolved techniques, capable of tracking optical fields and material responses on sub-cycle timescales, open a new frontier for probing strong coupling.[197] These approaches can directly observe light-matter energy exchange, dephasing processes and ultrafast population transfer between polariton branches, but their implementation is experimentally demanding due to the need for extreme stability and broadband sources. Theoretically, modelling field-resolved measurements requires time-dependent quantum electrodynamics frameworks able to describe electron dynamics, cavity fields and plasmonic response on attosecond-femtosecond scales. Progress in this direction will bridge the gap between semiclassical pictures and fully quantum descriptions, revealing the limits of coherence, the onset of nonlinearities and the true temporal structure of Rabi oscillations in realistic systems.

**4. Chemical reactivity, hot-carrier dynamics, and hybrid nanophotonics.** In plexcitonic systems, electromagnetic coupling coexists with chemical interactions and hot-carrier processes. Understanding how these channels interplay requires experimental setups capable of tracking charge-transfer events and nuclear motion on femtosecond timescales. Theoretically, incorporating non-adiabatic electronic-nuclear dynamics into hybrid quantum/classical models remains an outstanding challenge. Achieving this would bridge strong coupling and plasmonic photocatalysis, revealing whether hybrid states can steer chemical selectivity, modify reaction landscapes, or control hot-carrier injection pathways. This convergence could enable predictive design of polariton-assisted catalytic platforms.

**5. Chiral, collective, and many-molecule effects.** Strong coupling typically involves large ensembles of emitters, yet local inhomogeneities, molecular orientation, and nanoscale fields critically shape the resulting hybrid states. Experimentally, mapping these variations demands high-resolution spectroscopies and nanofabrication precision, while theory must capture collective effects in systems too large for fully quantum descriptions. Similar challenges arise in chiral or symmetry-broken platforms, where minute geometric or molecular asymmetries govern polaritonic optical activity. Coordinated effort will enable the development of collective-local hybrid models, clarifying how cooperative effects can be exploited to control energy flow, enantioselectivity, or nonlinear optical response.

**6. Toward unified multiscale models of realistic nanosystems.** Among the most overarching challenges, it remains the creation of theoretical tools capable of treating electronic structure, nuclear motion, plasmonic response, and cavity fields in a single self-consistent framework. Experimental advances now access coupling regimes in sub-nanometer cavities, open cavity geometries, and multimode metasurfaces, pushing the photophysics of polaritons beyond the validity of simplified models. Developing scalable, predictive multiscale approaches, validated against advanced ultrafast experiments, will be essential for developing design principles based on strong coupling for chemistry, photonics, and materials science.

## Author contributions

The manuscript was written through the contributions of all authors. All authors have given approval to the final version of the manuscript.

## Conflicts of interest

There are no conflicts to declare.

## Data availability

No new primary research data were generated or included as part of this Feature article.

## Acknowledgements

N.M., B. J. and A.K.P. acknowledge financial support from the Swedish Research Council (Grant No. 2021-05784 and No. 2025-04734), the Knut and Alice Wallenberg Foundation (Grant No. 2023.0089), the European Research Council (Grant No. 101116253), European Union’s Horizon 2020 Research and Innovation Programme under the Marie Skłodowska-Curie Actions (Grant No. 101205787), Kempestiftelserna (UCMR “Excellence by Choice” program Grant No. JCK-2130.3) and WISE-UmU/Kempe-program (Grant No. JCSMK 23-198).
M.M. acknowledges financial support from the ERC-StG ULYSSES grant agreement no. 101077181 funded by the European Union, and from the PhotoControl project under the PRIN 2022 MUR program (contract no. 2022N8PBLM, CUP: D53D23009190001). A.S. acknowledges the European Union’s Horizon Europe research and innovation programme under the Marie Skłodowska-Curie Action PATHWAYS HORIZON-MSCA-2023-PF-GF (grant agreement no. 101153856).
C.C. acknowledges financial support from the European Research Council (ERC) under the European Union’s Horizon 2020 research and innovation programme (grant agreement No. 818064) and MUR-FARE Ricerca in Italia: Framework per l’attrazione ed il rafforzamento delle eccellenze per la Ricerca in Italia - III edizione. Prot. R20YTA2BKZ.
M.V. acknowledges University of Milan for funding his postdoctoral fellowships “La bellezza degli aggregati: da nano a astro particelle”. F.B. thanks the financial support by MONSTER, a project of the Italian National Centre for HPC, Big Data and Quantum computing (ICSC, CUP B93C22000620006) funded by the European Union – NextGenerationEU, through PNNR. F.B and M.V thanks also the constructive discussion within the EPSRC grant (EP/W017075/1).